\documentclass{sig-alternate-05-2015}

\usepackage{booktabs}
\usepackage{amsmath}
\usepackage{amssymb}
\usepackage[normalem]{ulem} 
\usepackage{algorithmicx, algpseudocode}
\usepackage[linesnumbered,ruled,vlined]{algorithm2e}
\usepackage{graphicx}
\usepackage{subfig}
\usepackage{caption}
\usepackage{booktabs}
\usepackage{multirow}
\usepackage{epstopdf}
\usepackage{colortbl}
\usepackage{balance}
\usepackage{enumitem}

\newtheorem{theorem}{Theorem}
\newtheorem{corollary}{Corollary}

\newtheorem{Lemma}{Lemma}
\newtheorem{Definition}{Definition}

\newcommand\I{\mathbb{I}}
\newcommand\E{\mathbb{E}}

\newcommand\R{\mathcal{R}}

\newcommand{\Cov}{\mathrm{Cov}}

\def\IM{\textsf{IM}}

\def\PIMA{\textsf{PIMA}}
\def\OPT{\textsf{OPT}}

\def\RIS{\textsf{RIS}}
\def\PIMA{\textsf{SSA}}
\def\DPIMA{\textsf{D-SSA}}

\usepackage{mathtools}

\DeclarePairedDelimiter\floor{\lfloor}{\rfloor}

\SetKwComment{Comment}{$\triangleright$\ }{} 

\begin{document}

\makeatletter
\def\@copyrightspace{\relax}
\makeatother


%
\title{Stop-and-Stare: Optimal Sampling Algorithms for Viral Marketing in Billion-scale Networks}
%
%
%
%
%

\numberofauthors{3} 
%
\author{
%
%
\alignauthor
Hung T. Nguyen\\
\affaddr{CS Department}\\
\affaddr{Virginia Commonwealth Univ.}\\
\affaddr{Richmond, VA, USA}\\
\email{hungnt@vcu.edu}
\alignauthor
My T. Thai\\
\affaddr{CISE Department}\\
\affaddr{University of Florida}\\
\affaddr{Gainesville, Florida, USA}\\
\email{mythai@cise.ufl.edu}
\alignauthor
Thang N. Dinh\thanks{Corresponding author.}\\
\affaddr{CS Department}\\
\affaddr{Virginia Commonwealth Univ.}\\
\affaddr{Richmond, VA, USA}\\
\email{tndinh@vcu.edu}
}
\additionalauthors{Additional authors: John Smith (The Th{\o}rv{\"a}ld Group,
email: {\texttt{jsmith@affiliation.org}}) and Julius P.~Kumquat
(The Kumquat Consortium, email: {\texttt{jpkumquat@consortium.net}}).}
\date{30 July 1999}

\maketitle
\begin{abstract}
\textit{Influence Maximization} (\IM{}), that seeks a small set of key users who spread the influence widely into the network, is a core problem in multiple domains. It finds applications in viral marketing, epidemic control, and assessing cascading failures within complex systems. Despite the huge amount of effort, \IM{} in billion-scale networks such as Facebook, Twitter, and World Wide Web has not been satisfactorily solved. Even the state-of-the-art methods such as TIM+ and IMM may take days on those networks. 

In this paper, we propose \PIMA{} and \DPIMA{}, two novel sampling frameworks for \IM-based viral marketing problems. \PIMA{} and \DPIMA{} are up to 1200 times faster than the SIGMOD'15 best method, IMM, while providing the same $(1-1/e-\epsilon)$ approximation guarantee. Underlying our frameworks is an innovative \textit{Stop-and-Stare} strategy in which they \emph{stop at exponential check points} to verify (\emph{stare}) if there is adequate statistical evidence on the solution quality. Theoretically, we prove that \PIMA{} and \DPIMA{} are the first approximation algorithms that use  (asymptotically) minimum numbers of samples, meeting strict theoretical  thresholds characterized for IM. The absolute superiority of \PIMA{} and \DPIMA{} are confirmed through extensive experiments on real network data for \IM{} and another topic-aware viral marketing problem, named TVM.
\end{abstract}


\section{Introduction}
\label{sec:intro}

Viral Marketing, in which brand-awareness information is widely spread via the word-of-mouth effect, has emerged as one of the most effective marketing channels. It is becoming even more attractive with the explosion of social networking services such as Facebook\footnote{http://newsroom.fb.com/company-info/} with 1.5 billion monthly active users or Instagram\footnote{https://instagram.com/press/} with more than 3.5 billion daily like connections. To create a successful viral marketing campaign, one needs to seed the content with a set of individuals with high social networking influence. Finding such a set of users is known as the \textit{Influence Maximization} problem.

Given a network and a budget $k$, Influence Maximization (\IM{}) asks for $k$ influential users who can spread the influence widely into the network. Kempe et al. \cite{Kempe03} were the first to formulate \IM{} as a combinatorial optimization problem on the two pioneering diffusion models, namely,  \emph{Independent Cascade} (IC) and \emph{Linear Threshold} (LT). They prove \IM{} to be NP-hard and provide a natural greedy algorithm that yields  $(1 - 1/e - \epsilon)$-approximate solutions for any $\epsilon >0$. This celebrated work has motivated a vast amount of work on \IM{} in the past decade \cite{Leskovec07, Chen10, Goyal11, Goyal11_celf++,Cohen14, Ohsaka14, Tang14, Nguyen13_l, Dinh14, Shen12, Nguyen16}. However, most of the existing methods either too slow for billion-scale networks \cite{Kempe03,Leskovec07, Goyal11, Goyal11_celf++,Cohen14, Ohsaka14} or ad-hoc heuristics without performance guarantees \cite{Chen09_2,Chen10,Jung12,Wang10}.

The most scalable methods with performance guarantee for \IM{} are TIM/TIM+\cite{Tang14} and latter IMM\cite{Tang15}. They utilize a novel \RIS{} sampling technique introduced by Borgs et al. in \cite{Borgs14}. All these methods attempt to generate a $(1-1/e-\epsilon)$ approximate solution with minimal numbers of RIS samples. They use highly sophisticated estimating methods to make the number of RIS samples close to  some theoretical thresholds $\theta$ \cite{Tang14, Tang15}. However, they all share two shortcomings: 1) the number of generated samples can be arbitrarily larger than $\theta$, and 2) the thresholds $\theta$ are not shown to be the minimum among their kinds.

In this paper, we 1) unify the approaches in \cite{Borgs14, Tang14, Tang15} to characterize the necessary number of RIS samples to achieve $(1-1/e-\epsilon)$-approximation guarantee; 2) design two novel sampling algorithms \PIMA{} and \DPIMA{} aiming towards achieving minimum number of RIS samples. In the first part, we begin with defining \RIS{} framework which consists of two necessary conditions to achieve the $(1-1/e-\epsilon)$ factor and classes of \textit{\RIS{} thresholds} on the sufficient numbers of \RIS{} samples, generalizing $\theta$ thresholds in \cite{Tang14,Tang15}. The minimum threshold in each class is then termed \emph{type-1 minimum threshold}, and the minimum among all type-1 minimum thresholds is termed \emph{type-2 minimum threshold}.


In the second part, we develop the \emph{Stop-and-Stare Algorithm} (\PIMA{}) and its dynamic version \DPIMA{} that guarantee to achieve, within constant factors, the two minimum thresholds, respectively. Both \PIMA{} and \DPIMA{} follow the \emph{stop-and-stare} strategy which can be efficiently applied to many optimization problems over the samples and guarantee some constant times the minimum number of samples required. In short, the algorithms keep generating samples and \emph{stop} at \emph{exponential check points} to verify (\emph{stare}) if there is adequate statistical evidence on the solution quality for termination. This strategy will be shown to address both of the shortcomings in \cite{Tang14,Tang15}: 1) guarantee to be close to the theoretical thresholds and 2) the thresholds are minimal by definitions. The dynamic algorithm, \DPIMA{}, improves over \PIMA{} by automatically and dynamically selecting the best parameters for the \RIS{} framework. We note that the Stop-and-Stare strategy combined with \RIS{} framework enables \PIMA{} and \DPIMA{} to meet the minimum thresholds without explicitly computing/looking for these thresholds. That is in contrast to previous approaches \cite{Borgs14,Tang14,Tang15} which all find some explicit unreachable thresholds and then probe for them with unbounded or huge gaps.

Our experiments show that both \PIMA{} and \DPIMA{} outperform the best existing methods up to several orders of magnitudes w.r.t running time while returning comparable seed set quality. More specifically, on Friendster network with roughly 65.6 million nodes and 1.8 billion edges, \PIMA{} and \DPIMA{}, taking 3.5 seconds when $k=500$, are up to 1200 times faster than IMM. We also run CELF++ (the fastest greedy algorithm for \IM{} with guarantees) on Twitter network with $k=1000$ and observe that \DPIMA{} is $2\cdot10^9$ times faster. Our contributions are summarized as follows.
\begin{itemize}
	\item We generalize the \RIS{} sampling methods in \cite{Borgs14,Tang14,Tang15} into a general framework which characterizes the necessary conditions to guarantee the $(1-1/e-\epsilon)$-approximation factor. Based on the framework, we define classes of \RIS{} thresholds and two types of minimum thresholds, namely, type-1 and type-2.
	\item We propose the Stop-and-Stare Algorithm (\PIMA{}) and its dynamic version, \DPIMA{}, which both guarantee a $(1-1/e-\epsilon)$-approximate solution and are the first algorithms to achieve, within constant factors, the type-1 and type-2 minimum thresholds, respectively. Our proposed methods are not limited to solve influence maximization problem but also can be generalized for an important class of hard optimization problems over samples/sketches.
    \item Our framework and approaches are generic and can be applied in principle to sample-based optimization problems to design high-confidence approximation algorithm using (asymptotically) \emph{minimum number of samples}.
	\item We carry extensive experiments on various real networks with up to several billion edges to show the superiority in performance and comparable solution quality. To test the applicability of the proposed algorithms, we apply our methods on an \IM{}-application, namely, Targeted Viral Marketing (TVM). The results show that our algorithms are up to 1200 times faster than the current best method on \IM{} problem and, for TVM, the speedup is up to 500 times.
\end{itemize}

Note that this paper does not focus on distributed/parallel computation, however our algorithms are amenable to a distributed implementation which is one of our future works. 

\textbf{Related works.} Kempe et al.~\cite{Kempe03} formulated the influence maximization problem as an optimization problem. 
They show the problem to be NP-complete and devise an $(1-1/e - \epsilon)$ greedy algorithm.
Later, computing the exact influence is shown to be \#P-hard~\cite{Chen10}. 
Leskovec et al.~\cite{Leskovec07} study the influence propagation in a different perspective in which they aim to find a set of nodes in networks to detect the spread of virus as soon as possible. They improve the simple greedy method with the lazy-forward heuristic (CELF), which is originally proposed to optimize submodular functions in~\cite{Minoux78}, obtaining an (up to) 700-fold speedup.

Several heuristics are developed to find solutions in large networks. While those heuristics are often faster in practice, they fail to retain the $(1-1/e -\epsilon)$-approximation guarantee and produce lower quality seed sets. Chen et al. \cite{Chen09} obtain a speedup by using an influence estimation for the IC model. For the LT model, Chen et al. \cite{Chen10} propose to use local directed acyclic graphs (LDAG) to approximate the influence regions of nodes. In a complement direction, there are recent works on learning the parameters of influence propagation models \cite{Goyal10, Kutzkov13}. The influence maximization is also studied in other diffusion models including the majority threshold model \cite{Dinh14_pids} or when both positive and negative influences are considered \cite{Zhang13} and when the propagation terminates after a predefined time \cite{Dinh14_pids,Chen12}.
Recently, \IM{} across multiple OSNs have been studied in \cite{Shen12} and \cite{Du13} studies the \IM{} problem on continuous-time diffusion models.

Recently, Borgs et al. \cite{Borgs14} make a theoretical breakthrough and present an $O(kl^2 (m+n)\log^2 n/\epsilon^3)$ time algorithm for \IM{} under IC model. Their algorithm (RIS) returns a $(1-1/e-\epsilon)$-approximate solution with probability at least $1 - n^{-l}$. In practice, the proposed algorithm is, however, less than satisfactory due to the rather large hidden constants. In sequential works, Tang et al. \cite{Tang14, Tang15} reduce the running time to $O( (k+l) (m+n) \log n/\epsilon^2)$ and show that their algorithm is also very efficient in large networks with billions of edges. Nevertheless, Tang's algorithms have two weaknesses: 1) intractable estimation of maximum influence and 2) taking union bounds over all possible seed sets in order to guarantee a single returned set.

\textbf{Organization.} The rest of the paper is organized as follows: In Section~\ref{sec:model}, we introduce two fundamental models, i.e., LT and IC, and the \IM{} problem definition. We, subsequently, devise the unified \RIS{} framework, \RIS{} threshold and two types of \RIS{} minimum thresholds in Section~\ref{sec:ris}. Section~\ref{sec:alg} and~\ref{sec:approx} will present the \PIMA{} algorithm and prove the approximation factor as well as the achievement of type-1 minimum threshold. In Section~\ref{sec:extension}, we propose the dynamic algorithm, \DPIMA{} and prove the approximation together with type-2 minimum threshold property. Finally, we show experimental results in Section~\ref{sec:exp} and draw some conclusion in Section~\ref{sec:con}.

\section{Models and Problem Definition}
\label{sec:model}

This section will formally define two most essential propagation models, e.i., \emph{Linear Threshold} (LT) and \emph{Independent Cascade} (IC), that we consider in this work and followed by the problem statement of the Influence Maximization (\IM{}).

We abstract a network using a weighted graph $G=(V, E, w)$ with $|V|=n$ nodes and $|E|=m$ directed edges. Each edge $(u, v) \in E$ is associated with a weight $w(u, v) \in [0, 1]$ which indicates the probability that $u$ influences $v$. By convention, $w(u,v) = 0$ if $(u,v) \notin E$.

\subsection{Propagation Models}

In this paper, we study two fundamental diffusion models, namely, Linear Threshold (LT) and Independent Cascade (IC). Assume that we have a set of seed nodes $S$, the propagation processes under these two models happen in discrete rounds. At round $0$, all nodes in $S$ are active (influenced) and the others are inactive. In the subsequent rounds, the newly activated nodes will try to activate (or influence) their neighbors. Once a node $v$ becomes active, it will remain active till the end. The process stops when no more nodes get activated. The distinctions of the two models are described as follows:

\emph{Linear Threshold (LT) model}. The edge weights in LT model must satisfy the condition $\sum_{u \in V} w(u, v) \leq 1$. At the beginning of the propagation process, each node $v$ selects a random threshold $\lambda_v$ uniformly at random in range $[0, 1]$. In round $t\geq 1$, an inactive node $v$ becomes active if $\sum_{ \text{active } u \text{ at round } t-1} w(u, v) \geq \lambda_v$. Let $\I(S)$ denote the expected number of active nodes at the end of the diffusion process given the seed set $S$, where the expectation is taken over all $\lambda_v$ values from their uniform distribution. We call $\I(S)$ the \emph{influence spread} of $S$ under the LT model.

\emph{Independent Cascade (IC) model}. At round $t \ge 0$, when a node $u$ gets activated, initially or by another node, it has a single chance to activate each inactive neighbor $v$ with the successful probability proportional to the edge weight $w(u,v)$. An activated node remains active til the end of the diffusion process. For a set $S$ of nodes, we also denote $\I(S)$ as the influence spread of $S$ under the IC model, expected number of active nodes where the expectation is taken over the states of the random edges.

We summarize the frequently used notations in Table~\ref{tab:syms}.
\renewcommand{\arraystretch}{1.2}

\setlength\tabcolsep{3pt}
\begin{table}[hbt]\small
	\centering
	\caption{Table of notations}
	\vspace{-0.1in}
	\begin{tabular}{|p{1.5cm}|p{6.5cm}|}
		\addlinespace
		\toprule
		\bf Notation  &  \quad \quad \quad \bf Description \\
		\midrule 
		$n, m$ & \#nodes, \#edges of graph $G=(V, E, w)$.\\
		\hline
		$\I(S)$ & Influence Spread of seed set $S\subseteq V$.\\
		\hline
		$\OPT_k$ & The maximum $\I(S)$ for any size-$k$ seed set $S$.\\
		\hline
		$\hat S_k$ & The returned size-$k$ seed set of \PIMA{}/\DPIMA{}.\\
		\hline
		$S^*_k$ & An optimal size-$k$ seed set, i.e., $\I(S^*_k) = \OPT_k$.\\
		\hline
		$R_j$ & A random RR set.\\
		\hline
		$\R$ & A collection of random RR sets.\\
		\hline
		$\Cov_{\R}(S)$ & \#RR sets $R_j \in \R$ covered by $S$, i.e., $R_j \cap S \ne \emptyset$.\\           
		\hline
		$\hat \I_{\R}(S), \hat \I(S)$& $\frac{\Cov_{\R}(S)}{|\R|}$.\\           		
		\hline
		$\Upsilon(\epsilon,\delta)$ &
		$\Upsilon(\epsilon,\delta) = (2+\frac{2}{3}\epsilon)\ln \frac{1}{\delta} \frac{1}{\epsilon^2}$.\\
		\bottomrule
	\end{tabular}%
	\label{tab:syms}%
\end{table}%

\subsection{Problem Definition}
Given the propagation models defined previously, the Influence Maximization (IM) problem is defined as follows,
\begin{Definition}[Influence Maximization (\IM{})]Given a graph $G=(V,E,w)$, an integer $1 \leq k \leq |V|$ and a propagation model, the Influence Maximization problem asks for a seed set $\hat S_k \subset V$ of $k$ nodes that maximizes the influence spread $\I(\hat S_k)$ under the given propagation model.
\end{Definition}

\section{Unified RIS framework}
\label{sec:ris}
This section presents the unified \RIS{} framework, generalizing all the previous methods of using \RIS{} sampling \cite{Borgs14,Tang14,Tang15,Nguyen16} for \IM{}. The unified framework characterizes the sufficient conditions to guarantee an $(1-1/e-\epsilon)$-approximation in the framework. Subsequently, we will introduce the concept of \RIS{} threshold in terms of the number of necessary samples to guarantee the solution quality and two types of minimum \RIS{} thresholds, i.e., type-1 and type-2.
\subsection{Preliminaries}
\subsubsection{RIS sampling}
\label{subsec:ris}

The major bottle-neck in the traditional methods for \IM{} \cite{Kempe03, Leskovec07, Goyal11, Nguyen2013} is the inefficiency in estimating the influence spread. To address that, Borgs et al. \cite{Borgs14} introduced a novel sampling approach for \IM{}, called Reverse Influence Sampling (in short, \RIS{}), which is the foundation for TIM/TIM+\cite{Tang14} and IMM\cite{Tang15}, the state-of-the-art methods.
%
%
\begin{figure}[h!] \centering
	\includegraphics[width=\linewidth]{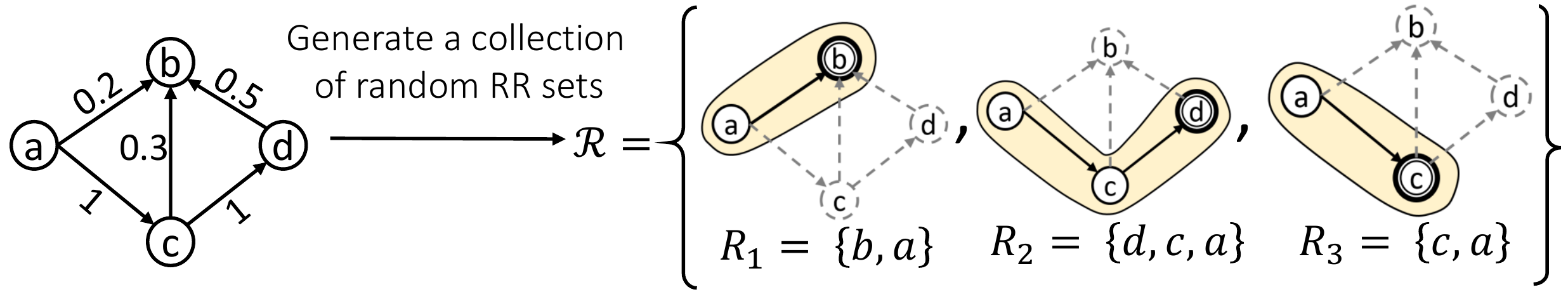}
	\caption{An example of generating random RR sets under the LT model. Three random RR sets $R_1, R_2$ and $R_3$ are generated. Node $a$ has the highest influence and is also the most frequent element across the RR sets.}
	\label{fig:rr}
\end{figure}

Given a graph $G=(V, E, w)$, \RIS{} captures the influence landscape of $G$ through generating a set $\R$ of random \textit{Reverse Reachable (RR) sets}. The term `RR set' is also used in TIM/TIM+ \cite{Tang14,Tang15} and referred to as `hyperedge' in \cite{Borgs14}. Each RR set $R_j$ is a subset of $V$ and constructed as follows,
\begin{Definition}[Reverse Reachable (RR) set]
	Given $G=(V, E, w)$, a random RR set $R_j$ is generated from $G$ by 1) selecting a random node $v \in V$ 2) generating a sample graph $g$ from $G$ and 3) returning $R_j$ as the set of nodes that can reach $v$ in $g$.
\end{Definition}
Node $v$ in the above definition is called the \emph{source} of $R_j$. Observe that $R_j$ contains the nodes that can influence its source $v$. 

If we generate multiple random RR sets, influential nodes will likely appear frequently in the RR sets. Thus a seed set $S$ that \emph{covers} most of the RR sets will likely maximize the influence spread $\I(S)$. Here a seed set $S$ covers an RR set $R_j$, if $S \cap R_j \neq \emptyset$. For convenience, we denote the coverage of set $S$ as follows,
\begin{align}
	\Cov_{\R}(S) = \sum_{R_j \in \R} \min\{|S \cap R_j|,1\}
\end{align}

An illustration of this intuition and how to generate RR sets is given in Fig.~\ref{fig:rr}. In the figure, three random RR sets are generated following the LT model with sources $b$, $d$ and $c$, respectively. The influence of node $a$ is the highest among all the nodes in the original graph and also is the most frequent node across the RR sets. This observation is captured in the following lemma in \cite{Borgs14}.	

\begin{Lemma}[\cite{Borgs14}]
	\label{lem:Borgs}
	Given $G=(V, E, w)$, a seed set $S\subset V$, for a random RR set $R_j$ generated from $G$
	\begin{align}
	\I(S) = n \Pr[S \text{ covers } R_j].
	\end{align}
\end{Lemma}

Lemma~\ref{lem:Borgs} says that the influence of a node set $S$ is proportional to the probability that $S$ intersects with a random RR set. Define
\[
	\hat \I_{\R}(S) = \frac{\Cov_{\R}(S)}{|\R|},
\]
an approximate of $\I(S)$. When the context is clear, we also ignore $\R$ and write $\hat \I(S)$ instead of $\hat \I_{\R}(S)$. Thus, to find $S$ that maximize $\I(S)$ we can find $S$ to maximize $\hat I(S)$, i.e., to find subset $S$ that covers as many $R_j$ as possible. The most important question addressed in this paper is about the \emph{minimum size of $\R$ to provide bounded-error guarantees}.

\subsubsection{$(\epsilon,\delta)$-approximation}

The bounded-error guarantee we seek for in our influence maximization algorithms, $(1-1/e-\epsilon)$ with probability at least $(1-\delta)$, is based on the concept of $(\epsilon, \delta)$-approximation. 
\begin{Definition}[$(\epsilon, \delta)$-approximation]
	Let $Z_1, Z_2,...$ be i.i.d. random variables in $[0, 1]$ with mean $\mu_Z$ and variance $\sigma_Z^2$. A \textit{Monte Carlo estimator}
	\begin{align}
	\hat \mu_Z = \frac{1}{T}\sum_{i=1}^{T}Z_i
	\end{align}
	is said to be an $(\epsilon, \delta)$-approximation of $\mu_Z$ if
	\begin{align}
	\label{eq:ed_approx}
	\Pr [(1-\epsilon)\mu_Z \leq \hat \mu_Z \leq (1 + \epsilon) \mu_Z] \geq 1 - \delta
	\end{align}
\end{Definition}


Let $R_1, R_2, R_3,\ldots, R_j,\ldots$ be the random RR sets generated in either \PIMA{} or \DPIMA{} algorithms.  Given a subset of nodes $S \subset V$, define   $Z_j = \min\{|R_j \cap S|,1\}$, the Bernouli random variable with mean $\E[Z_j]=\I(S)/n$. Further, define $Y_j = Z_j - \E[Z_j]$, then 
$Y_j$ is a \emph{martingale} \cite{Tang15}, i.e., $\E[Y_i | Y_1, Y_2,\ldots,Y_{i-1}] = Y_{i-1}$ and $\E[Y_i]<+\infty$.  This martingale view of $Y_j$ is adopted from \cite{Tang15} to cope with the fact that  $Y_j$ might be weakly dependent due to the stopping condition. Let $\hat \mu_Z = \frac{1}{T}\sum_{i=1}^{T}Z_i$, an estimation of $\mu_Z$. We use the same concentration inequalities from Corollaries 1 and 2 in \cite{Tang15}.

\begin{Lemma}[\cite{Tang15}]
	\label{lem:chernoff}
 For $T > 0$ and $\epsilon > 0$, the following inequalities hold,
	\vspace{-0.05in}
	\begin{align}
		\label{eq:plus}
		\Pr[\hat \mu >  (1+ \epsilon) \mu]  &\leq
		\exp{(\frac{-T\mu\epsilon^2}{2 + \frac{2}{3}\epsilon})},\\
		\label{eq:minus}
		\Pr[\hat \mu <  (1- \epsilon) \mu] &\leq \exp{(\frac{-T\mu\epsilon^2}{2})}.
	\end{align}
\end{Lemma}

Equivalently, we can derive from Lem.~\ref{lem:chernoff} the sufficient number of samples to provide an $(\epsilon, \delta)$-approximation.

\begin{corollary}
	For fixed $\epsilon >0$ and $ \delta \in (0, 1)$,
	\begin{align}
		\label{col:opt}
&\Pr[\hat \mu >  (1+ \epsilon) \mu]  \leq  \delta, \text{if }  T \geq  \frac{2+\frac{2}{3}\epsilon}{\epsilon^2} \ln \frac{1}{\delta}\frac{1}{\mu}=\Upsilon(\epsilon, \delta) \frac{1}{\mu},\\
&\Pr[\hat \mu <  (1- \epsilon) \mu] \leq \delta, \text{if } T \geq \frac{2}{\epsilon^2}\ln (\frac{1}{\delta})\frac{1}{\mu}.
	\end{align}
\end{corollary}

\subsection{RIS Framework and Thresholds}
\label{sub:ris_frame}
Based on Lem.~\ref{lem:Borgs}, the \IM{} problem can be solved by the following two-step algorithm.
\begin{itemize}
	\item Generate a collection of RR sets, $\R$, from $G$.
	\vspace{-0.05in}
	\item Use the greedy algorithm for the Max-coverage problem \cite{Khuller1999} to find a seed set $\hat S_k$ that covers the maximum number of RR sets and return $\hat S_k$ as the solution.
\end{itemize}

As mentioned, the core issue is to determine the minimum $\theta(\epsilon,\delta)$ given a predefined setting of $\epsilon,\delta$. For \IM{}, this means \emph{``How many RR sets are sufficient to provide a good approximate solution?''}. \cite{Tang14,Tang15} propose two such theoretical thresholds and two probing techniques to realistically estimate those thresholds. However, their thresholds are not known to be any kind of minimum and the probing method is \textit{ad hoc} in \cite{Tang14} or far from the proposed threshold in \cite{Tang15}. Thus, they cannot provide any guarantee on the optimality of the number of samples generated.


We look into the cores of the techniques in \cite{Tang14,Tang15,Borgs14,Nguyen16} and capture the essential conditions to achieve an $(1-1/e-\epsilon)$ approximation for Influence Maximization problem. By satisfying these critical conditions, we aim to achieve a better approach rather than the prescribing a explicit threshold $\theta$ as in previous work \cite{Tang14,Tang15,Borgs14,Nguyen16}.

\textbf{RIS Critical conditions.} Suppose that there is an optimal seed set $S^*_k$, which has the maximum influence in the network\footnote{If there are multiple optimal sets with influence, $\OPT_k$, we choose the first one alphabetically to be $S^*_k$.}. Given $0 \leq \epsilon,\delta \leq 1$, our unified \RIS{} framework enforces two conditions:
\begin{align}
	\label{eq:sk}
	\Pr[\hat \I(\hat S_k) \leq (1 + \epsilon_a) \I(\hat S_k)] \geq 1 - \delta_a
\end{align}
\vspace{-0.2in}

\noindent and
\vspace{-0.05in}
\begin{align}
	\label{eq:sk*}
	\Pr[\hat \I(S^*_k) \geq (1 - \epsilon_b)\OPT_k] \geq 1 - \delta_b
\end{align}
\noindent where $\delta_a + \delta_b \leq \delta\text{ and } (1-\frac{1}{e})\frac{\epsilon_a+ \epsilon_b}{1+\epsilon_a} \leq \epsilon$.

Based on the above conditions, we define the \RIS{} threshold as the following.
\begin{Definition}[\RIS{} Threshold]
	\label{def:type1}
	Given a graph $G$, $\epsilon_a \in (0, \infty)$ and $\epsilon_b, \delta_a, \delta_b \in (0, 1)$, $N(\epsilon_a, \epsilon_b, \delta_a, \delta_b)$ is called an \RIS{} Threshold in $G$ w.r.t $\epsilon_a, \epsilon_b, \delta_a, \delta_b$,  if  $|\R| \geq N(\epsilon_a, \epsilon_b, \delta_a, \delta_b)$ implies  Eqs.~\ref{eq:sk} and~\ref{eq:sk*} hold together.
\end{Definition}
The \RIS{} threshold gives a \emph{sufficient condition} to achieve a $(1-1/e-\epsilon)$-approximation as stated below.
\begin{theorem}
	\label{theo:type1}
	Given a graph $G$, $\epsilon_a \in [0, \infty)$, and $\epsilon_b, \delta_a, \delta_b \in (0, 1)$, let $\epsilon = (1-\frac{1}{e})\frac{\epsilon_a+ \epsilon_b}{1+\epsilon_a}$ and $\delta \geq \delta_a + \delta_b$, if the number of RR sets $|\R| \geq N(\epsilon_a, \epsilon_b, \delta_a, \delta_b)$, then the two-step algorithm in our \RIS{} framework returns $\hat S_k$ satisfying
	\begin{align}
		\Pr[\I(\hat S_k) \geq (1 - 1/e -\epsilon) \emph{\OPT}_k] \geq 1 - \delta.
	\end{align}
That is $\hat S_k$ is an $(1-1/e-\epsilon)$-approximate solution with high probability (w.h.p.)
\end{theorem}
\noindent \textbf{Existing \RIS{} thresholds.}
For any $\epsilon, \delta \in (0, 1)$, Tang et al. established in \cite{Tang14} an \RIS{} threshold,
\begin{align}
	\label{eq:theta_1}
	N(\frac{\epsilon}{2},\frac{\epsilon}{2},\frac{\delta}{2}(1- \tbinom{n}{k}^{-1}),\frac{\delta}{2}\tbinom{n}{k}) = (8 + 2\epsilon)n \frac{ \ln 2/\delta + \ln {n \choose k} }{ \epsilon^2 \OPT_k}
\end{align}
In a later study \cite{Tang15}, they reduced this number to another \RIS{} threshold from Theorem~1 in \cite{Tang15},
\begin{align}
	\label{eq:theta}
	N(\epsilon_1,\epsilon-\epsilon_1,\frac{\delta}{2}(1-\tbinom{n}{k}^{-1}),\frac{\delta}{2}\tbinom{n}{k}^{-1}) = 2n \frac{((1-1/e)\alpha + \beta)^2}{ \epsilon^2 \OPT_k},
\end{align}
where $\alpha = (\ln\frac{2}{\delta})^{\frac{1}{2}}$, $\beta =~ (1-1/e)^{\frac{1}{2}}(\ln \frac{2}{\delta} + \ln \tbinom{n}{ k})^{\frac{1}{2}}$ and $\epsilon_1 =\frac{\epsilon\cdot \alpha}{(1-1/e)\alpha + \beta}$.

Simplify the above equation, we have
\begin{align*}
&((1-1/e)\alpha + \beta)^2 \leq 2( (1-1/e)^2 \alpha^2+ \beta^2) \\
&= 2(1-1/e) ((1-1/e)\ln\frac{2}{\delta} + \ln\frac{2}{\delta}+\ln \tbinom{n}{k} ) \\
&\leq 2(1-1/e) (2\ln\frac{2}{\delta}+\ln \tbinom{n}{k}).
\end{align*}
Thus, we obtain a simplified threshold,
\begin{align}
	\label{eq:theta_2}
	N & = 4(1-\frac{1}{e})n \frac{2\ln (2/\delta) + \ln {n \choose k}}{\epsilon^2 \OPT_k }\\
	 &\leq 8(1-\frac{1}{e})\frac{\ln (2/\delta) + \ln {n \choose k}}{\epsilon^2 } \frac{n}{k}
\end{align}

Unfortunately, computing $\OPT_k$ is intractable, thus, the proposed algorithms have to generate $\theta \frac{\OPT_k}{KPT^+}$ RR sets, where $KPT^+$ is the expected influence of a node set obtained by sampling k nodes with replacement from $G$ and the ratio $\frac{\OPT_k}{KPT^+} \geq 1$ is not upper-bounded. That is they may generate many times more RR sets than needed as in \cite{Tang14}.

\subsection{Two Types of Minimum Thresholds}
Based on the definition of \RIS{} threshold, we now define two strong theoretical limits, i.e. type-1 minimum and type-2 minimum thresholds. In Section \ref{sec:approx}, we will prove that our first proposed algorithm, \PIMA{}, achieves, within a constant factor, a type-1 minimum threshold and later, in Section \ref{sec:extension}, our dynamic algorithm, \DPIMA{}, is shown to obtain, within a constant factor, the strongest type-2 minimum threshold.

If $N(\epsilon_a,\epsilon_b,\delta_a,\delta_b)$ is an \RIS{} threshold, then any $N$ such that $N \geq N(\epsilon_a,\epsilon_b,\delta_a,\delta_b)$ is also an \RIS{} threshold. We choose the smallest number over all such \RIS{} thresholds to be type-1 minimum as defined in Def.~\ref{def:type1_min}.
\begin{Definition}[type-1 minimum threshold]
	\label{def:type1_min}
	Given\\ $0 \leq \epsilon,\delta \leq 1$ and $\epsilon_a \in (0, \infty)$, $\epsilon_b, \delta_a, \delta_b \in (0, 1)$ satisfying $\delta_a + \delta_b \leq \delta\text{ and } (1-\frac{1}{e})\frac{\epsilon_a + \epsilon_b}{1+\epsilon_a} \leq \epsilon$, $N^{(1)}_{min}(\epsilon_a,\epsilon_b,\delta_a,\delta_b)$ is called a type-1 minimum threshold w.r.t $\epsilon_a, \epsilon_b,\delta_a,\delta_b$ if $N^{(1)}_{min}(\epsilon_a,\epsilon_b,\delta_a,\delta_b)$ is the smallest number of RR sets that satisfies both Eq.~\ref{eq:sk} and Eq.~\ref{eq:sk*}.
\end{Definition}

 All the previous methods \cite{Borgs14,Tang14,Tang15} try to approximate $N^{(1)}_{min}(\epsilon_a,\epsilon_b,\delta_a,\delta_b)$ for some setting of $\epsilon_a,\epsilon_b,\delta_a,\delta_b$, however, they fail to provide any guarantee on how close their numbers are to that threshold. In contrast, we show that \PIMA{} achieves, within a constant factor, a type-1 minimum threshold in Section~\ref{sec:approx}. 
Next, we give the definition of a stronger type-2 minimum threshold which is achieved by \DPIMA{} as shown in Section~\ref{sec:extension}.


\begin{Definition}[type-2 minimum threshold]
	\label{def:type2}
	Given\\ $0 \leq \epsilon,\delta \leq 1$, $N^{(2)}_{min}(\epsilon, \delta)$ is called the type-2 minimum threshold if
	\begin{align}
	N^{(2)}_{min}(\epsilon,\delta) = \min_{\epsilon_a,\epsilon_b,\delta_a,\delta_b}N^{(1)}_{min}(\epsilon_a,\epsilon_b,\delta_a,\delta_b)
	\end{align}
	where $(1-\frac{1}{e})\frac{\epsilon_a + \epsilon_b}{1+\epsilon_a}=\epsilon$ and $\delta_a + \delta_b = \delta$ and $\epsilon_a \in (0, \infty)$, $\epsilon_b, \delta_a, \delta_b \in (0, 1)$.
	\label{def:type2_min}
\end{Definition}
Type-2 minimum threshold is the tightest threshold that one can achieve using the \RIS-framework.

\section{Stop-and-Stare Algorithm (SSA)}
\label{sec:alg}
In this section, we present  Stop-and-Stare Algorithm (\PIMA{}), the first approximation algorithm that meets (asymptotically) a type-1 minimum threshold. 
\begin{algorithm}[!h]
	\caption{\PIMA{} Algorithm}
	\label{alg:main}
	\KwIn{Graph $G$, $0 \leq \epsilon, \delta \leq 1$, and a budget $k$}
	\KwOut{An $(1 - 1/e - \epsilon)$-optimal solution, $\hat S_k$ with at least $(1-\delta)$-probability}
	Choose $\epsilon_1,\epsilon_2,\epsilon_3$ satisfying Eqs. \ref{eq:epsilon}. For example, recommended values for $\epsilon_1, \epsilon_2, \epsilon_3$ are in Eq.~\ref{eq:set_ep}\\
	$N_{max} = 8\frac{1-1/e}{2+2\epsilon /3}\Upsilon\left(\epsilon, \frac{\delta}{6}/{n\choose k}\right)\frac{n}{k}$; 	$i_{max} = \lceil \log_2 \frac{2N_{max}}{\Upsilon(\epsilon, \delta/3)} \rceil$; \\
	$\Lambda = \Upsilon(\epsilon, \frac{\delta}{3 i_{max}}); \Lambda_1 \leftarrow (1 + \epsilon_1)(1 + \epsilon_2)\Upsilon(\epsilon_3,\frac{\delta}{3i_{max}})$\\
	$\R \leftarrow$ Generate $\Lambda$ random RR sets\\
	\Repeat{$|\R|\geq N_{max}$}{	
		Double the size of $\R$ with new random RR sets\\	
		$<$$\hat S_k,\hat \I(\hat S_k)$$>\leftarrow$ \textsf{Max-Coverage}($\R$, $k$, $n$)\\
		\If(\Comment*[f]{Condition C1}){$\Cov_{\R}(\hat S_k) \geq \Lambda_1$}{
			$\delta'_2 = \frac{\delta_2}{3i_{max} }; T_{max} = 2 |\R| \frac{1+\epsilon_2}{1-\epsilon_2} \frac{\epsilon_3^2}{\epsilon_2^2}$\\
			$\I_c(\hat S_k) \leftarrow $ \textsf{Estimate-Inf}($G, \hat S_k, \epsilon_2, \delta'_2$, $T_{max}$)\\
			\If(\Comment*[f]{Condition C2}){$\hat \I(\hat S_k) \leq (1+\epsilon_1) \I_c(\hat S_k)$}{
				\textbf{return} $\hat S_k$
			}
		}
	}
	\textbf{return} $\hat S_k$
\end{algorithm}

\subsection{SSA Algorithm}
\label{subsec:main}
\setlength{\textfloatsep}{5pt}
At a high level, \PIMA{}, presented in Alg.~\ref{alg:main}, consists of multiple iterations. In each iteration, it follows the \RIS{} framework to generate (additional) RR sets and uses the \textsf{Max-Coverage} (Alg.~\ref{alg:cov}) to find a candidate solution $\hat S_k$. If $\hat S_k$ passes the quality check, Lines 8-12, the algorithm stops and outputs $\hat S_k$. Otherwise, it doubles the number of RR sets and advances to the next iteration. The name Stop-and-Stare is based on the view that the algorithm ``scans'' through a stream of samples and \emph{stops} at exponential check points to \emph{stare} at the the generated samples to see if it can find a provably good solution. We enforce a nominal cap on the number of samples	$N_{max} = 8\frac{1-1/e}{2+2\epsilon /3}\Upsilon\left(\epsilon, \frac{\delta}{6}/{n\choose k}\right)\frac{n}{k}$. Thus, the  number of iterations is at most $i_{max} = \lceil \log_2 \frac{2N_{max}}{\Upsilon(\epsilon, \delta/3)} \rceil=O(\log_2 n)$ (Line 2).

Specifically, the algorithm starts by determining parameters $\epsilon_1, \epsilon_2, \epsilon_3$ satisfying $(1-\frac{1}{e})\frac{\epsilon_a + \epsilon_b}{1+\epsilon_a}=\epsilon$ (Line 1). For each iteration $t=1,2,\ldots, i_{max}$, \PIMA{} doubles the number of generated RR sets in $\mathcal R$. Thus, the number of samples at an iteration $t$ is $|\mathcal R| = \Lambda 2^{t-1}$, where $
\Lambda  =\Upsilon(\epsilon,\delta/(3i_{max}))$.
After that, \PIMA{} invokes \textsf{Max-Coverage} (Alg.~\ref{alg:cov}) to find a candidate solution $\hat S_k$ and its influence estimation \[
\hat \I(\hat S_k) = \frac{\Cov_{\mathcal R} (\hat S_k)n}{|\mathcal R|}.
\]
The condition $\Cov_{\R}(\hat S_k) \geq \Lambda_1$ (Line 8) is to guarantee that there are sufficient samples to estimate the influence accurately within a relative error $\epsilon_3$. If the condition is met, \PIMA{} independently generates another collection of RR sets $\mathcal R'$ in \textsf{Estimate-Inf} (Alg.~\ref{alg:check}) to obtain an accurate estimation of $\hat S_k$ influence (with a relative error $\epsilon_2$).  This estimation is compared against $\hat \I(\hat S_k)$ and the \PIMA{} stops when the two estimations are close (Line 11), i.e., when 
\[
\hat \I(\hat S_k) \leq (1+\epsilon_1) \I_c(\hat S_k).
\]

\textbf{Stopping conditions}. Ignore the rare case that \PIMA{} reaches the cap $N_{max}$ on the number of samples. \PIMA{} stops only when the following two \textit{stopping conditions} are met.
\begin{itemize}
	\item[(C1)] The 1st condition $\Cov_{\R}(\hat S_k) \geq \Lambda_1$ (Line 8) ensures that the  influence of  $S^*_k$ can be estimated with a relative error at most $\epsilon_3$ as shown in Lems. \ref{lem:com1} and~\ref{lem:com2}. 
	\item[(C2)] The 2nd condition $\hat \I(\hat S_k) \leq (1+\epsilon_1) \I_c(\hat S_k)$ (Line 11) guarantee that the estimation $\hat \I(\hat S_k)$ is not far from the error-bounded estimation $\I_c(\hat S_k)$ returned by the \textsf{Estimate-Inf} procedure. Recall that $\I_c(\hat S_k)$  has a relative error at most $\epsilon_2$ comparing to the true influence $\I(\hat S_k)$. 
\end{itemize}

\noindent As we will prove in Sec.~\ref{sec:approx}, the two stopping conditions are sufficient to guarantee the $(1-1/e-\epsilon)$-approximation of $\hat S_k$.

\begin{algorithm}
	\caption{\textsf{Max-Coverage} procedure}
	\label{alg:cov}
	\KwIn{RR sets ($\R$), $k$ and number of nodes ($n$)}
	\KwOut{An $(1 - 1/e)$-optimal solution, $\hat S_k$ and its estimated influence $\I_c(\hat S_k)$}
	$\hat S_k = \emptyset$\\
	\For{$i=1:k$}{
		$\hat v \leftarrow \arg \max_{\{v\in V\}}(\Cov_{\R}(\hat S_k\cup \{v\}) - \Cov_{\R}(\hat S_k))$\\
		Add $\hat v$ to $\hat S_k$\\
	}
	\textbf{return} $<$$\hat S_k, \Cov_{\R}(\hat S_k)\cdot n/|\R|$$>$
\end{algorithm}

\textbf{Finding Max-Coverage}. Standard greedy algorithm in \textsf{Max-coverage} is used to find $\hat S_k$. The algorithm repeatedly selects node $u$ with maximum marginal gain, the number of RR sets that are covered by $u$ but not the previously selected nodes. The well-known result in \cite{Nemhauser81} states that $\Cov_{\mathcal R}(\hat S_k)$ is at leat $(1-1/e)$ the maximum coverage obtained by any size-$k$ seed set. This algorithm can be implemented in linear time in terms of the total size of  the RR sets \cite{Borgs14}.

\textbf{Influence Estimation}.
\label{subsec:inf_ver}
\textsf{Estimate-Inf}, presented in Alg.~\ref{alg:check}, gives an estimation $\I_c(S)$ with one-side error guarantee
\[
\Pr[ \I_c(S) \leq (1+\epsilon') \I(S) ] \geq 1 - \delta'.
\]
\setlength{\textfloatsep}{5pt}
\begin{algorithm}
	\caption{\textsf{Estimate-Inf} procedure}
	\label{alg:check}
	\KwIn{A seed set $S \subset V$, $\epsilon'>0$, $\delta' \in (0, 1)$ and maximum number of samples, $T_{max}$}
	\KwOut{$\I_c(S)$ or $-1$ if exceeds $T_{max}$ samples.}
	$\Lambda_2 = 1+(1+\epsilon') \Upsilon(\epsilon', \delta')$\\
	$\Cov = 0$\\
	\For{$T = 1:T_{max}$}{
		Generate $R_j \leftarrow $\RIS{}$(G)$\\
		$\Cov = \Cov + \min\{|R_j \cap S|,1\}$\\
		\If{$\Cov \ge \Lambda_2$}{
			\textbf{return} $n\Lambda_2/T$ \tcp*{$n$: number of nodes}
		}
	}
	\textbf{return }-1 \tcp*{Exceeding $T_{max}$ RR sets}
\end{algorithm}
 The algorithm generates RR sets $R_j$ and counts the number of ``successes'', defined as the number of RR sets that intersect with $S$. When the number of successes  reaches $\Lambda_2 =1 +(1+\epsilon')\Upsilon(\epsilon', \delta')$, the algorithm returns $\I_c(S)=\frac{\Lambda_2 n}{T}$, where $T$ is the number of generated RR sets. 
 
\textsf{Estimate-Inf} is based on the Stopping-Rule algorithm in \cite{Dagum00} with an important difference. The algorithm stops and return $-1$ if $T_{max}$ samples has been generated. Choosing $T_{max}$ proportional to the number of samples in $\mathcal R$ (Line 9, \PIMA) avoid time-wasting on estimating influence for $\hat S_k$ at early iterations in \PIMA.  Those early $\hat S_k$ candidates often have small influence, thus, require up to $\Omega(n)$ samples to estimate. Without the cap $T_{max}$, \PIMA{} will suffer a quadratic (or worse) time complexity.
 
Similar to the proof of the stopping theorem in \cite{Dagum00}, we obtain the following lemma with the proof in the appendix.
\begin{Lemma}
	\label{cor:check}
	When \textsf{Estimate-Inf} terminates within $T_{max}$ samples, the returned estimation $\I_c(S)$  satisfies
	\begin{align}
	\Pr[\I_c(S) \leq (1 + \epsilon')\I(S)] & \geq 1 - \delta'.
	\end{align}
\end{Lemma}

In \PIMA{} Lines 9 and 10, \textsf{Estimate-Inf} is invoked with the parameters $\epsilon'=\epsilon_2$, $\delta' = \delta_2/(3i_{max})$, and $T_{max} = \Theta(|\mathcal R|)$.

\subsection{Parameter Settings for \PIMA}
\label{subsec:para_set}
In \PIMA{}, we can select arbitrary $\epsilon_1, \epsilon_2,\epsilon_3 \in (0, 1)$ as long as they satisfy
\begin{align}
\label{eq:epsilon}
(1-\frac{1}{e})\frac{\epsilon_1+\epsilon_2+\epsilon_1\epsilon_2+\epsilon_3}{(1+\epsilon_1)(1+\epsilon_2)} \leq \epsilon\end{align}
In practice, the selection of $\epsilon_1, \epsilon_2$ and $\epsilon_3$ has considerate effect on the running time. Through our experiments, we observe good performance yields when 
\begin{itemize}
\item $\epsilon_1 > \epsilon \approx \epsilon_3$\ \ for small networks
\item $\epsilon_1 \approx \epsilon \approx \epsilon_3$\ \ \ for moderate network (few million edges)
\item $\epsilon_1 \ll \epsilon_2 \approx \epsilon_3$ for large networks (hundreds of millions of edges).
\end{itemize}
%
%

For simplicity, we use the following \emph{default setting} for \PIMA.
\begin{align}
\label{eq:set_ep}
\epsilon_2  &= \epsilon_3 = (1-1/e)^{-1} \epsilon/2\\
\epsilon_1  &= \frac{1+(1-1/e-\epsilon)^{-1} \epsilon/2}{1+\epsilon_2} - 1.
\end{align}
For example, when $\epsilon = 0.1$ we can set 
\begin{align}
\epsilon_1 = 1/78, \epsilon_2 = \epsilon_3 = 2/25.
\end{align}

In Sect.~\ref{sec:extension}, we will later propose \DPIMA{}, a Stop-and-Stare algorithm with  ``dynamic'' parameters. \DPIMA{} can automatically select a near-optimal setting of $\epsilon_1, \epsilon_2, \epsilon_3$.

\section{SSA Theoretical Analysis}
\label{sec:approx}

In this section, we will prove that \PIMA{} returns a $(1 - 1/e - \epsilon)$-approximate solution w.h.p. in Subsec.~\ref{subsec:approx}. Subsequently, \PIMA{} is shown to require no more than a constant factor of a type-1 minimum threshold of RR sets w.h.p. in Subsec.~\ref{subsec:pima_rr}.

\subsection{Approximation Guarantee}
\label{subsec:approx}
We will prove that \PIMA{} returns a $(1-1/e-\epsilon)$-approximate solution $\hat S_k$  w.h.p.
 The major portion of the proof is  to bound the probabilities of the following three bad events
 \begin{enumerate}
  \item $|R| \geq N_{max}$ and $\I( \hat S_k) < (1-1/e - \epsilon)$ 
  \item The error in the estimation $\I_c(\hat S_k)$ exceeds $\epsilon_2$ (Lem.~\ref{lem:com1})
  \item The error in the estimation $\hat \I(S^*_k)$, the estimation of the  $OPT_k$, exceeds $\epsilon_3$ (Lem.~\ref{lem:com2}). 
 \end{enumerate}
Finally, Theorem~\ref{theo:approx}, assuming none of the bad events happen,  shows that $\I(\hat S_k) \geq (1 - 1/e - \epsilon) \OPT_k.$
  
The probability of the first bad event follows directly from the threshold $\theta$ in Eq.~\ref{eq:theta_2} with $\delta$ replaced by $\delta/3$.
\begin{Lemma}
\label{lem:caps} We have
\[
\Pr[ |\R| \geq N_{max} \text{ and } \I(\hat S_k) < (1-1/e-\epsilon)\emph{\OPT}_k ] \leq \delta/3.
\]
\end{Lemma}


Since, we do not know the iteration that \PIMA{} will stop, we will bound the probabilities of the other two bad events for all iterations.  The bound on the relative error of $\I_c(\hat S_k)$:  
\begin{Lemma}
	\label{lem:com1}
	For any iteration  $t=1,2,\ldots,i_{max}$  in \PIMA, 
	\begin{align}
		\Pr [\I_c(\hat S_k) > (1 + \epsilon_{2})\I(\hat S_k)] \leq \delta/{3i_{max}}.
	\end{align}
\end{Lemma}
\begin{proof}
		The inequality holds trivially if \textsf{Estimate-Inf} return $-1$. Otherwise, it follows from Lem.~\ref{cor:check} with $\epsilon'=\epsilon_2$, $\delta' = \delta_2/(3i_{max})$.
\end{proof}

Since $|\R| =\Lambda 2^{t-1}$ is fixed, we apply the Chernoff's bound in Lem.~\ref{lem:chernoff} over $|\R|$ random variables
to obtain the following error bound on the estimation of $\hat \I(S^*_k)$. 
\begin{Lemma}
	\label{lem:com2}
	For any iteration $i=1,2,\ldots,i_{max}$ in \PIMA, 
	\begin{align}
		\Pr [& \hat \I(S^*_k) < (1 - \epsilon^{(i)}_3)\emph{\OPT}_k] \leq  \delta/(3i_{max})
	\end{align}
	where $\epsilon^{(i)}_3 = \sqrt{\frac{2n \ln \frac{3 i_{max}}{\delta}}{|\R| \emph{\OPT}_k}}$, and $|\R| =\Lambda 2^{t-1}$ at iteration $i$.
\end{Lemma}

Lem.~\ref{lem:com1} and \ref{lem:com2} are sufficient to prove the approximation guarantee of \PIMA{} as stated by the following theorem.
\begin{theorem}
	\label{theo:approx}
	Given $0 \leq \epsilon, \delta \leq 1$, \PIMA{} returns a seed set $\hat S_k$ satisfying
	\begin{align}
		\Pr[\I(\hat S_k) \geq (1 - 1/e - \epsilon) \emph{\OPT}_k] \geq 1-\delta.
	\end{align}
\end{theorem}

\subsection{Achieving Type-1 Minimum Threshold}
\label{subsec:pima_rr}
We will show that for any $\epsilon_a, \epsilon_b, \delta_a, \delta_b$ satisfying the conditions of \RIS{} threshold (Def.~\ref{def:type1}), there exists a setting of $\epsilon_1, \epsilon_2, \epsilon_3$ such that \PIMA{} stops within $O(N^{(1)}_{min}(\epsilon_a, \epsilon_b, \delta_a, \delta_b) )$ samples (w.h.p.)


We need to bound the total number of RR sets generated by \PIMA{}. Recall that \PIMA{} generates two different types of RR sets:  1) RR sets in $\R$ to find $\hat S_k$ through solving \textsf{Max-Coverage} and 2) RR sets in \textsf{Estimate-Inf} for the stopping condition C2. At each iteration, the number of type 2 RR sets is at most $2\frac{1 + \epsilon_2}{1-\epsilon_2}\frac{\epsilon_3^2}{\epsilon_2^2} |\R| = \Theta(|\R|)$.  Thus, the core part is to prove that:  \emph{``\PIMA{} will stop w.h.p. when $|\R| = O( N_1(\epsilon_a, \epsilon_b, \delta_a, \delta_b))$''}.

\textbf{Our assumptions.} Under the assumptions that make the Chernoff's bound (Lem.~\ref{lem:chernoff}) tight up to a constant in the exponent, we show that \PIMA{} stops within $O(N^{(1)}_{min}(\epsilon_a, \epsilon_b, \delta_a, \delta_b) )$. The assumptions, referred to as the \emph{\underline{range conditions}}, are as follows.
\begin{itemize}
\item $\OPT_k \leq \frac{1}{2} |V|$. That is no $k$ nodes can influence more than half of the nodes in the network. This assumption guarantees $\mu \leq 1/2$, needed for the tightness of Chernoff's bound in Lem.~\ref{lem:tightness}.
\item $\epsilon \leq 1/4$. The constant $1/4$ can be replaced by any constant $c < 1$, assuming $\delta$ is sufficiently small. This assumption guarantees that $\epsilon_b \leq 1/2$, which is also needed for  Lem.~\ref{lem:tightness}.
\item $1/\delta =\Omega(n)$. This assumption guarantee that $\delta$ is sufficiently small (Lem.~\ref{lem:tightness}). This is compatible with the settings in the previous works \cite{Tang14,Tang15,Nguyen16}, in which $\delta = 1/n$. 
\end{itemize}

Consider   positive $\epsilon_a, \epsilon_b, \delta_a, \delta_b \in (0, 1)$ satisfying 
\begin{align}
\label{eq:risp1}
(1-\frac{1}{e})\frac{\epsilon_a+ \epsilon_b}{1+\epsilon_a}= \epsilon \leq \frac{1}{4} \text{ and }\\
\label{eq:risp2}
\delta_a + \delta_b = \delta <\frac{1}{\log_2 n}.
\end{align}
 We will determine suitable parameters $\epsilon_1, \epsilon_2, \epsilon_3$ for \PIMA{} so that
\begin{align}
\label{eq:epima}
(1-\frac{1}{e})\frac{\epsilon_1+\epsilon_2+\epsilon_1\epsilon_2+\epsilon_3}{(1+\epsilon_1)(1+\epsilon_2)} = \epsilon.
\end{align}

\noindent 	Denote  $T_1 = N^{(1)}_{min}(\epsilon_a, \epsilon_b, \delta_a, \delta_b)$. From Def.~\ref{def:type1_min} of the type-1 threshold,   $|\R|\geq T_1$ leads to
\begin{align}
\label{eq:eq11}
&\Pr [\hat \I_{\R}(\hat S_k) > (1 + \epsilon_a)\I(\hat S_k)] \leq \delta_a \text{ and }\\
\label{eq:eq22}
&\Pr [\hat \I_{\R}(S^*_k) < (1 - \epsilon_b)\OPT_k] \leq \delta_b.
\end{align}

An upper bound on the number of RR sets needed in $\R$ is given in the following lemma.
\begin{Lemma}
	\label{lem:tssa}
	Let $\epsilon_0 = \min\{ \epsilon_2, \epsilon_3, \epsilon_b\}$, and 
	\[
	T_{\PIMA} = \max\{ T_1, \alpha \Upsilon(\epsilon_0, \frac{\delta}{3i_{max}}) \frac{n}{\OPT_k} \},
	\]
	for some constant $\alpha>1$. Under the range conditions, 
	\[
	 	T_{\PIMA} = O( T_1).
	\]	
\end{Lemma}
%

Now we bound the estimation error in the \textsf{Estimate-Inf} procedure at each iteration. At  iteration $1\leq i\leq i_{max}$, 
\begin{align}
T_{max} = 2|\R|\frac{1+\epsilon_2}{1-\epsilon_2} \frac{\epsilon_3^2}{\epsilon_2^2} = 2^{i}\Lambda\frac{1+\epsilon_2}{1-\epsilon_2} \frac{\epsilon_3^2}{\epsilon_2^2}
\end{align}
is a fixed number. Denote by $\R_c$, the set of RR sets generated in \textsf{Estimate-Inf}. 
Apply the concentration inequality in Eq. (\ref{eq:plus}), for $T_{max}$ RR sets in $\R_c$ we  have
\begin{Lemma}
	\label{lem:eit}
	For iteration $1\leq i \leq i_{max}$ in \PIMA{}, let $\epsilon_2^{(i)} = \sqrt{ \frac{(\ln 1/\delta +\ln 3i_{max}) n}{ T_{max} }}$. The following holds
	\[
	\Pr[ (|\R_c| \geq T_{max}) \text{ and } \hat \I_{\R_c}(\hat S_k) < (1-\epsilon_2^{(i)}) \I(\hat S_k)] \leq \frac{\delta}{3i_{max}}.
	\]
\end{Lemma}

\begin{theorem}
	\label{theo:efficiencyssa}
Consider $\epsilon_a, \epsilon_b, \delta_a, \delta_b$ satisfying Eqs. (\ref{eq:risp1}) and (\ref{eq:risp2}). Under the range conditions, there exist \PIMA{} parameters  $\epsilon_1, \epsilon_2, \epsilon_3$, satisfying Eq. (\ref{eq:epima}), and a constant $c > 1$ such that if $|\R| \geq c N^{(1)}_{min}(\epsilon_a, \epsilon_b, \delta_a, \delta_b)$,  \PIMA{} will stop w.h.p.
\end{theorem}

Similarly, we can show the reverse direction.
\begin{theorem}
Consider \PIMA's with $\epsilon_1, \epsilon_1, \epsilon_2, \epsilon_3$, satisfying Eq. (\ref{eq:epima}) and $\epsilon_2 \leq \frac{\epsilon_1}{1+\epsilon_1}$. Under the range conditions, there exist  $\epsilon_a, \epsilon_b, \delta_a, \delta_b$ satisfying Eqs. (\ref{eq:risp1}) and (\ref{eq:risp2}),  and a constant $c > 1$ such that if $|\R| \geq c N_1(\epsilon_a, \epsilon_b, \delta_a, \delta_b)$,  \PIMA{} will stop w.h.p.
\end{theorem}
The proof is similar to that of Theorem \ref{theo:efficiencyssa} and is omitted.

\textbf{\PIMA{} Limitation.} First, the performance of \PIMA{} depends on the selection of the parameters $\epsilon_1, \epsilon_2, \epsilon_3$. While the presetting in Eq. \ref{eq:set_ep} provides decent performance for most cases, there will  be certain input that results in less than ideal performance. Secondly,  the samples in $\R'$, the sample pool to verify the quality of the candidate solution $\hat S_k$ are not used efficiently. They are only used once and then discarded. Alternative strategies that reuse the sample in $\R'$ may potentially reduce the number of the generated samples and provide better performance.


\section{Dynamic Stop-and-Stare Algo.}
\label{sec:extension}
In this section, we present \DPIMA{}, a stop-and-stare algorithm that automatically selects near-optimal $\epsilon_1, \epsilon_2, \epsilon_3$ settings. That is the sample size of \DPIMA{} meets, asymptotically, the type-2 minimum threshold, the strongest guarantee for methods  following the \RIS{} framework.

\begin{algorithm}[!ht]
	\caption{\DPIMA{} Algorithm}
	\label{alg:dpima}
	\KwIn{Graph $G$, $0 \leq \epsilon, \delta \leq 1$, and $k$}
	\KwOut{An $(1 - 1/e - \epsilon)$-optimal solution, $\hat S_k$}
	$N_{max} = 8\frac{1-1/e}{2+2\epsilon /3}\Upsilon\left(\epsilon, \frac{\delta}{6}/{n\choose k}\right)\frac{n}{k}$; \\
	$t_{max} = \lceil \log_2 (2 N_{max}/ \Upsilon(\epsilon, \frac{\delta}{3})) \rceil$; $t = 0$;\\
	$\Lambda = \Upsilon(\epsilon, \frac{\delta}{3 t_{max}}); \Lambda_1 = 1+(1+\epsilon) \Upsilon(\epsilon, \frac{\delta}{3 t_{max}})$; \\
	\Repeat{$|\R_t|\geq N_{max}$}{
		$t \leftarrow t + 1$; \\
		$\R_t = \{ R_1, \dots, R_{\Lambda2^{t-1}} \}$; \\
		$\R^{c}_t = \{ R_{\Lambda2^{t-1}+1}, \dots, R_{\Lambda 2^t} \}$; \\
		$<\hat S_k, \hat \I_t(\hat S_k)> \leftarrow \textsf{Max-Coverage}(\R_t, k)$;\\
		\If(\Comment*[f]{Condition D1}){$\Cov_{\R^{c}_t}(\hat S_k) \geq \Lambda_1  
		$}  
		{
			$\I^{c}_t(\hat S_k) \leftarrow \Cov_{\R^{c}_t}(\hat S_{k})\cdot n/|\R^{c}_t|$\\
			$\epsilon_1 \leftarrow \hat \I_t(\hat S_k)/\I^{c}_t(\hat S_k) - 1$\\
			$\epsilon_2 \leftarrow \epsilon \sqrt{\frac{n(1+\epsilon)}{2^{t-1} \I^{c}_t (\hat S_k)}}$;\\
			$\epsilon_3 \leftarrow \epsilon \sqrt{\frac{n(1+\epsilon)(1-1/e-\epsilon)}{(1+\epsilon/3)2^{t-1} \I^{c}_t (\hat S_k)}}$\\		
			$\epsilon_t = (\epsilon_1 + \epsilon_2 + \epsilon_1 \epsilon_2)(1-1/e-\epsilon) + (1-\frac{1}{e})\epsilon_3$\\
			\If(\Comment*[f]{Condition D2}){$\epsilon_t \leq \epsilon$ } 
				{\textbf{return} $\hat S_k$}
		}
	}
	\textbf{return} $\hat S_k$;
\end{algorithm}

%
The algorithm \DPIMA{}, summarized in Alg.~\ref{alg:dpima},
works on a \emph{single stream} of RR sets $R_1, R_2,...,R_i,...$. The algorithm consists of multiple iterations $t=1, 2,\ldots,t_{max}$, where $t_{max} = O(\log n)$ is the maximum number of iterations. 

At an iteration $t$, the algorithm looks into the first $\Lambda \times 2^{t}$ RR sets, for a fixed $\Lambda$ (Line~3), and divide those samples into two halves.
\begin{itemize}
\item 	The first half $
\R_t = \{ R_1, \dots, R_{\Lambda2^{t-1}} \}
$
will be used to find the candidate solution $\hat S_k$ via solving a max-coverage problem $\textsf{Max-Coverage}(\R_t, k)$.\\
\item 	The second half	
$
\R^{c}_t = \{ R_{\Lambda2^{t-1}+1}, \dots, R_{\Lambda 2^t} \}
$
will be used to verify the quality of the candidate solution $\hat S_k$.
\end{itemize}

Note that $\R_{t+1} = \R_t \cup \R^{c}_t$, thus, the samples used in verifying $\hat S_k$ will be reused to find the candidate solution in next iteration.

To verify whether $\hat S_k$ meets the approximation guarantee with high probability (whp), \DPIMA{}, in Line 9, will first apply the stopping rule condition  in \cite{Dagum00} to check if the number of samples in $\R^{c}_t$ are sufficient to guarantee an $(\epsilon, \frac{\delta}{3t_{max}})$-approximation of $\I(\hat S_k)$. If not, it advances to the next iteration. Otherwise, it will automatically estimate the best possible precision parameters $\epsilon_1, \epsilon_2, \epsilon_3$ in Lines 11 and 12. Once the combination of those precision parameter is sufficiently small, i.e., 
$$\epsilon_t = (\epsilon_1 + \epsilon_2 + \epsilon_1 \epsilon_2) (1-1/e-\epsilon) + (1-1/e)\epsilon_3 \leq \epsilon,$$ 
the algorithm returns $\hat S_k$ as an $(1-1/e-\epsilon)$-approximation solution (whp).

In the unfortunate event that the algorithm does not meet the condition $\epsilon_t \leq \epsilon$ for any $t$, it will terminate when the number of samples in the algorithm reaches to the cap $N_{max} = 8\frac{1-1/e}{2+2\epsilon /3} \Upsilon\left(\epsilon, \frac{1}{6}\delta/{n\choose k}\right)\frac{n}{k}$.

\subsection{Theoretical Guarantees Analysis}

We will subsequently show that \DPIMA{} achieves the $(1-1/e-\epsilon)$-approximation factor (whp) in Subsec.~\ref{subsub:dima_approx} and requires only, to within a constant factor, the strongest type-2 minimum threshold of the RR sets (whp) in Subsec.~\ref{subsub:dima_rr}.

\subsubsection{Approximation Guarantee}
\label{subsub:dima_approx}

We will show that \DPIMA{} returns a $(1-1/e-\epsilon)$ solution with probability at least $1-\delta$. For clarity, we present most of the proofs in the appendix.

Recall that \DPIMA{} stops when either 1)the number of samples exceeds the cap, i.e., $|\R_t| \geq N_{max}$ or 2) $\epsilon_t \leq \epsilon$ for some $t\geq 1$. In the first case,  $N_{max}$ were chosen to guarantee that 
$\hat S_k$ will be a $(1-1/e-\epsilon)$-approximation solution w.h.p.
\begin{Lemma}
\label{lem:cap}
Let $B^{(1)}$ be the bad event that 
\[
B^{(1)} = (|\R_t| \geq N_{max}) \cap (\I(\hat S_k) < (1-1/e-\epsilon)\emph{\OPT}_k).
\] We have
\[
	\Pr[B^{(1)}] \leq  \delta/3.
\]
\end{Lemma}

In the second case, the algorithm stops when $\epsilon_t \leq \epsilon$ for some
$1\leq t \leq t_{max}$. The maximum number of iterations $t_{max}$ is bounded by $O(\log n)$ as stated below.
\begin{Lemma}
\label{lem:tmax}
The number of iterations in \DPIMA{} is at most $t_{max} = O(\log n)$.
\end{Lemma}


For each iteration $t$, we will bound the probabilities of the bad events that lead to inaccurate estimations of  $\I(\hat S_k)$ through $\R^{c}_t$, and $\I(S^*_k)$ through $\R_t$(Lines~9 and~12).

\begin{Lemma}
\label{lem:bad2}
For each $1\leq t \leq t_{max}$, let 
\[
\hat \epsilon_t \text{ be the unique root of } f(x)=\frac{\delta}{3t_{max}},
\]
where $f(x)=\exp{\left(-\frac{N_t \frac{\I(\hat S_k)}{n} x^2 }{2+2/3x}\right)}$, and 
\[
\epsilon_t^*=
\epsilon \sqrt{\frac{n}{(1+\epsilon/3)2^{t-1} \emph{\OPT}_k}}.
\]
Consider the following bad events
\begin{align*}
B_t^{(2)} &= \left(
 \hat \I_t^{(c)}(\hat S_k) > (1+\hat \epsilon_t)\I(\hat S_k) \right),
\\
B_t^{(3)} &= \left(
 \hat \I_t(S^*_k) < (1-\epsilon_t^*) \emph{\OPT}_k]
 \right).
\end{align*}
We have 
\[
 \Pr[B_t^{(2)}],  \Pr[B_t^{(3)}] \leq \frac{\delta}{3t_{max}}.
\]
\end{Lemma}
%
%

\begin{Lemma}
\label{lem:e2e}
Assume that none of the bad events $B^{(1)}$, $B^{(2)}_t$, $B^{(3)}_t$  ($t =1..t_{max}$) happen and \DPIMA{} stops with some $\epsilon_t \leq \epsilon$. With $\hat \epsilon_t$ defined in Lem.~\ref{lem:bad2}, we have 
\begin{align}
	&\hat \epsilon_t < \epsilon \text{ and consequently }\\
	&\I^{(c)}_t(\hat S_k) \leq  (1+\hat \epsilon_t)\I(\hat S_k) \leq (1+\epsilon)\I(\hat S_k)
\end{align}
\end{Lemma}

We now prove the approximation guarantee of \DPIMA{}.

\begin{theorem}
	\label{theo:accuracydssa}
	\DPIMA{} returns an $(1-1/e-\epsilon)$-approximate solution with probability at least $(1-\delta)$.
\end{theorem}

\subsubsection{Achieving the Type-2 Minimum Threshold}
\label{subsub:dima_rr}
  Denote by $T_2=N^{(2)}_{min}(\epsilon,\delta)$, the type-2 minimum threshold defined in Def.~\ref{def:type2}. 	Under the \emph{range conditions}, we will prove that  \DPIMA{} meets the Type-2 minimum threshold, i.e., it requires $O( T_2)$ samples w.h.p. This is the strongest efficiency guarantee for algorithms following the \RIS{} framework.
	
	The proof is based on the observation that there must exist $\epsilon^*_a, \epsilon^*_b, \delta^*_a, \delta^*_b$ that 
		$N^{(1)}_{min}(\epsilon^*_a, \epsilon^*_b, \delta^*_a, \delta^*_b)=T_2$. Further,  within $O(T_2)$ we will have
		 $\epsilon_2, \epsilon_3 \leq \epsilon_b^*/3$ and $\epsilon_1 \approx \epsilon_a^*$. Then both conditions D1
		  ($\Cov_{\R^c_t}(\hat S_k) \geq \Lambda_1$)  and D2 ($\epsilon_t \leq \epsilon$) will be met and the algorithm will stop w.h.p.

\begin{theorem}
	\label{theo:dima}
	Given $\epsilon, \delta$, assume the \emph{range conditions} \DPIMA{} will stop w.h.p within $O(N^{(2)}_{min}(\epsilon,\delta))$ samples.
\end{theorem}

\begin{figure*}[!ht]
	\subfloat[NetHEPT]{
		\includegraphics[width=0.24\linewidth]{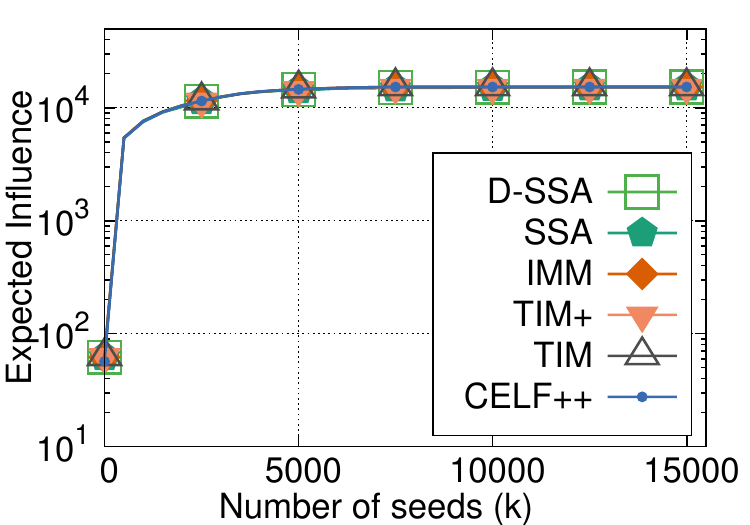}
	}
	\subfloat[NetPHY]{
		\includegraphics[width=0.24\linewidth]{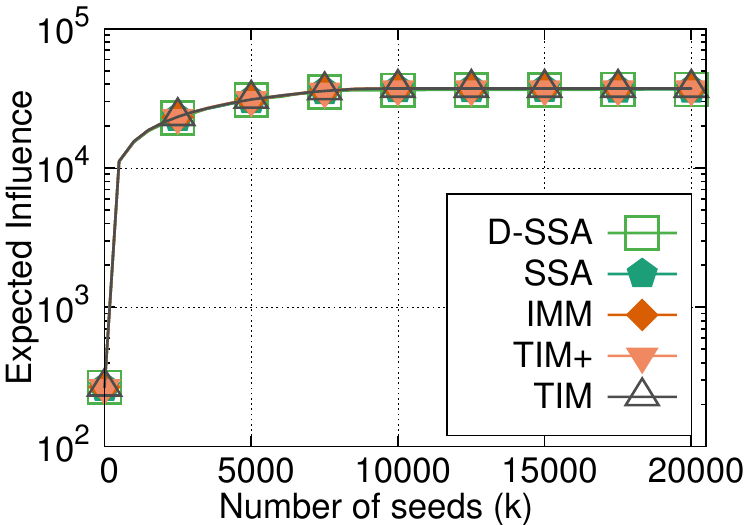}
	}
	\subfloat[DBLP]{
		\includegraphics[width=0.24\linewidth]{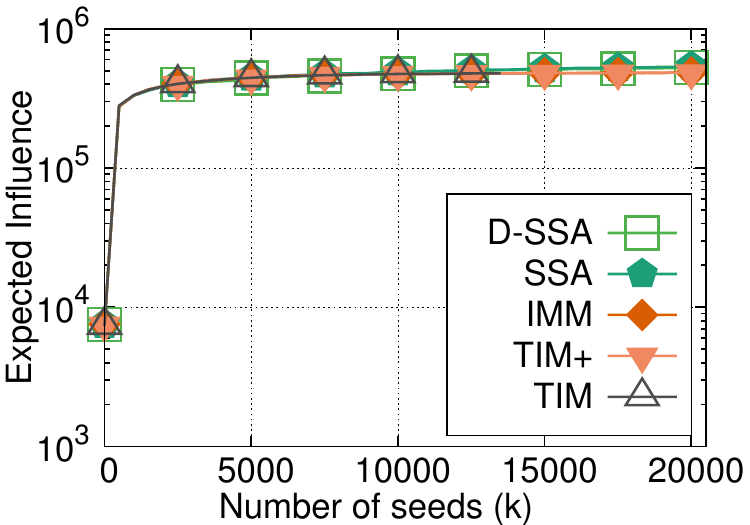}
	}
	\subfloat[Twitter]{
		\includegraphics[width=0.24\linewidth]{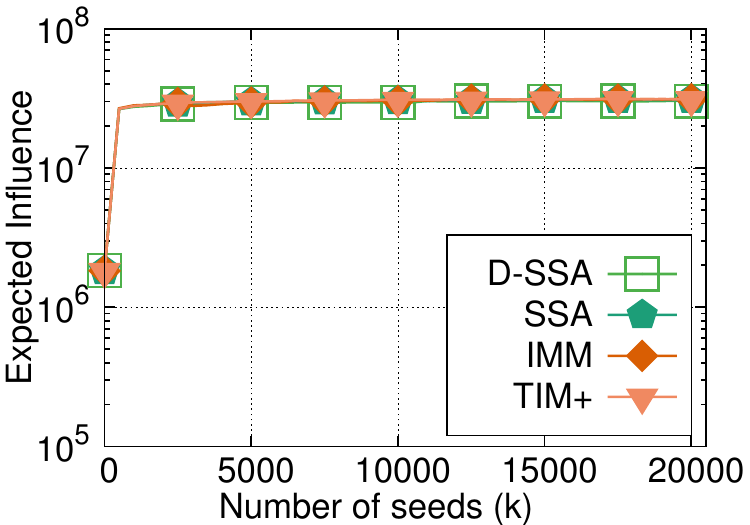}
	}
	\vspace{-0.1in}
	\caption{Expected Influence under LT model.}
	\label{fig:qual_inf_lt}
	\vspace{-0.2in}
\end{figure*}
\begin{figure*}[!ht]
	\subfloat[NetHEPT]{
		\includegraphics[width=0.24\linewidth]{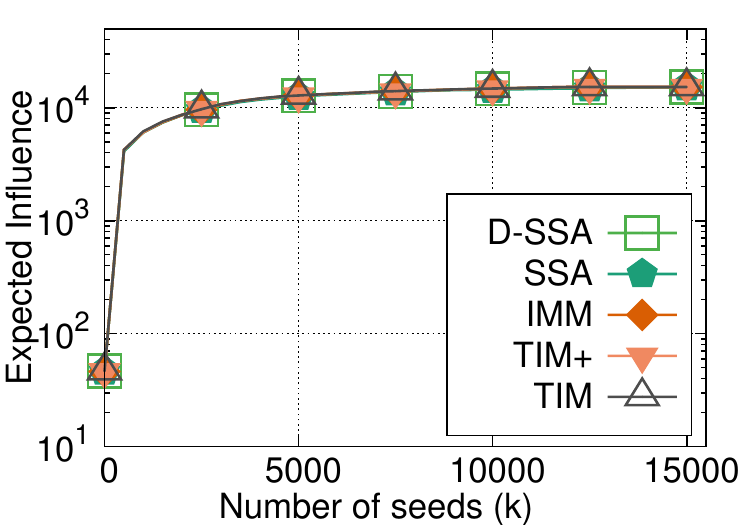}
	}
	\subfloat[NetPHY]{
		\includegraphics[width=0.24\linewidth]{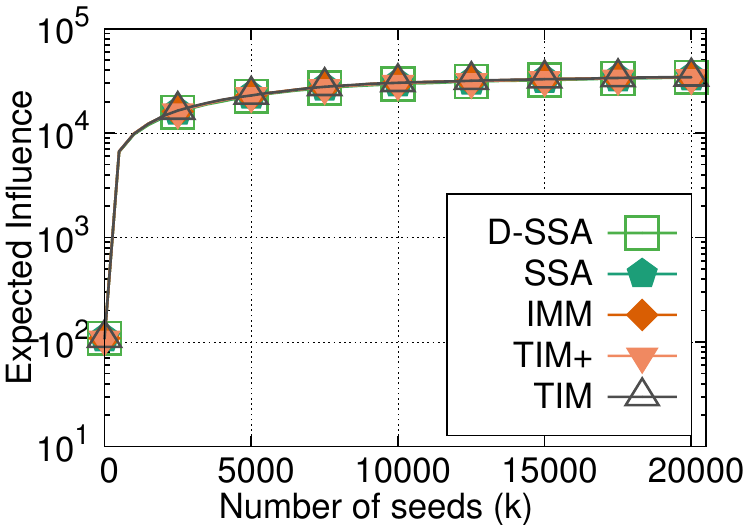}
	}
	\subfloat[DBLP]{
		\includegraphics[width=0.24\linewidth]{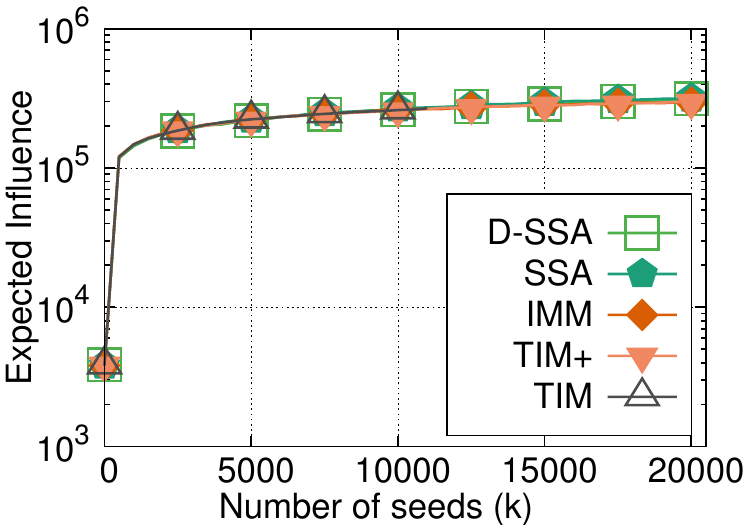}
	}
	\subfloat[Twitter]{
		\includegraphics[width=0.24\linewidth]{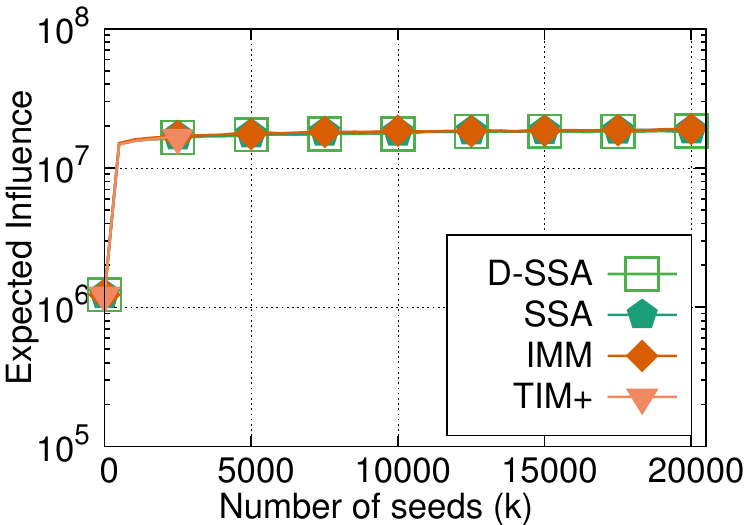}
	}
	\vspace{-0.1in}
	\caption{Expected Influence under IC model.}
	\label{fig:qual_inf_ic}
	\vspace{-0.2in}
\end{figure*}
\section{Experiments}
\label{sec:exp}
Backing by the strong theoretical results, we will experimentally show that \PIMA{} and \DPIMA{} outperform the existing state-of-the-art \IM{} methods by a large margin. Specifically, \PIMA{} and \DPIMA{} are several orders of magnitudes faster than IMM and TIM+, the best existing \IM{} methods with approximation guarantee, while having the same level of solution quality. \PIMA{} and \DPIMA{} also require several times less memory than the other algorithms. To demonstrate the applicability of the proposed algorithms, we apply our methods on a critical application of \IM{}, i.e., Targeted Viral Marketing (TVM) introduced in \cite{Li15} and show the significant improvements in terms of performance over the existing methods.
\vspace{-0.15in}
\setlength\tabcolsep{3pt}
\begin{table}[!htb]
	\caption{Datasets' Statistics}
	\vspace{-0.05in}
	\label{tab:data_sum}
	\centering
	\begin{tabular}{ l  r  r  c  r}\toprule
		\textbf{Dataset} & \bf \#Nodes& \bf \#Edges & \bf Avg. degree\\\midrule
		NetHELP\footnote{From http://snap.stanford.edu}& 15K & 59K & 4.1\\
		NetPHY\footnotemark[\value{footnote}] & 37K & 181K & 13.4\\
		Enron\footnotemark[\value{footnote}] & 37K & 184K & 5.0\\
		Epinions\footnotemark[\value{footnote}] & 132K & 841K & 13.4\\
		DBLP\footnotemark[\value{footnote}] & 655K & 2M & 6.1\\
		Orkut\footnotemark[\value{footnote}] & 3M & 234M & 78\\
		Twitter \cite{Kwak10} & 41.7M & 1.5G & 70.5\\
		Friendster\footnotemark[\value{footnote}] & 65.6M & \textbf{3.6G} & 54.8\\\bottomrule
		\hline
	\end{tabular}
\end{table}

\subsection{Experimental Settings}
All the experiments are run on a Linux machine with 2.2Ghz Xeon 8 core processor and 100GB of RAM. We carry experiments under both LT and IC models on the following algorithms and datasets.

\textbf{Algorithms compared.} On \IM{} experiments, we compare \PIMA{} and \DPIMA{} with the group of top algorithms that provide the same $(1-1/e-\epsilon)$-approximation guarantee. More specifically, CELF++ \cite{Goyal112}, one of the fastest greedy algorithms, and IMM \cite{Tang15}, TIM/TIM+ \cite{Tang14}, the best current \RIS{}-based algorithms, are selected. For experimenting with TVM problem, we apply our Stop-and-Stare algorithms on this context and compare with the most efficient method for the problem, termed KB-TIM, in \cite{Li15}.


\textbf{Datasets.}
For experimental purposes, we choose a set of 8 datasets from various disciplines: NetHEPT, NetPHY, DBLP are citation networks, Email-Enron is communication network, Epinions, Orkut, Twitter and Friendster are online social networks. The description summary of those datasets is in Table \ref{tab:data_sum}. 
On Twitter network, we also have the actual tweet/retweet dataset and we use these data to extract the target users whose tweets/retweets are relevant to a certain set of keywords. The experiments on TVM are run on the Twitter network with the extracted targeted groups of users.

\textbf{Remark.} Since Orkut and Friendster are \textbf{undirected} networks, within those networks we \emph{replace each edge by two oppositely directed edges} (arcs). This contrasts to the conference version of this paper in which the Orkut and Friendster networks are treated as directed networks.

\textbf{Parameter Settings.} For computing the edge weights, we follow the conventional computation as in \cite{Tang14,Chen09_2,Goyal11,Nguyen2013}, the weight of the edge $(u,v)$ is calculated as $w(u,v) = \frac{1}{d_{in}(v)}$ where $d_{in}(v)$ denotes the in-degree of node $v$.



In all the experiments, we keep $\epsilon=0.1$ and $\delta = 1/n$ as a general setting or explicitly stated otherwise. For the other parameters defined for particular algorithms, we take the recommended values in the corresponding papers if available. We also limit the running time of each algorithm in a run to be within 24 hours.
\begin{figure*}[!ht]
	\subfloat[NetHEPT]{
		\includegraphics[width=0.24\linewidth]{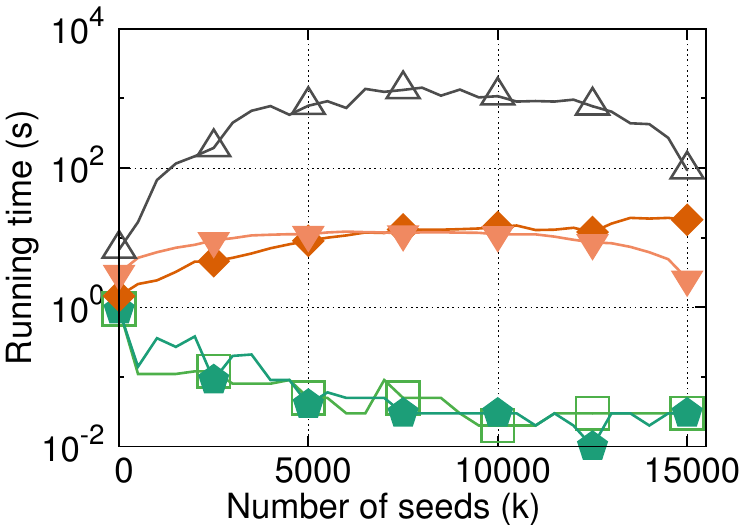}
	}
	\subfloat[NetPHY]{
		\includegraphics[width=0.24\linewidth]{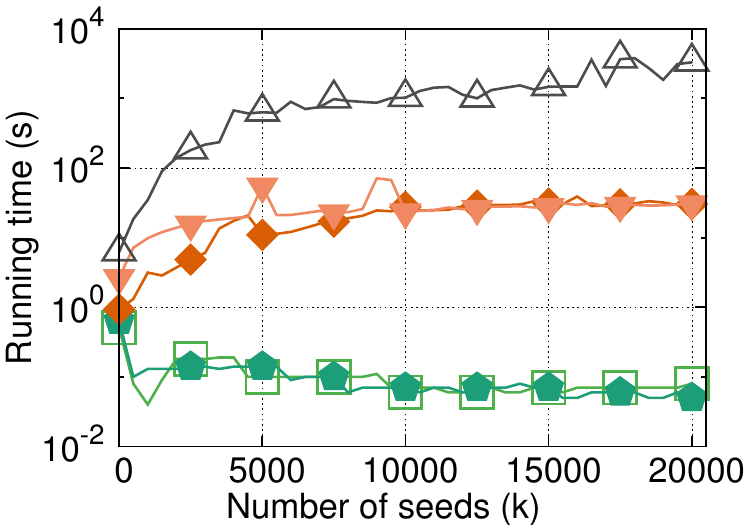}
	}
	\subfloat[DBLP]{
		\includegraphics[width=0.24\linewidth]{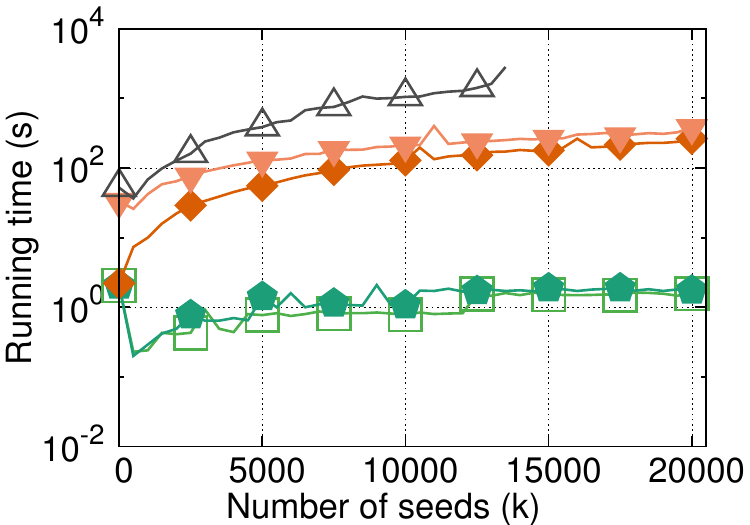}
	}
	\subfloat[Twitter]{
		\includegraphics[width=0.24\linewidth]{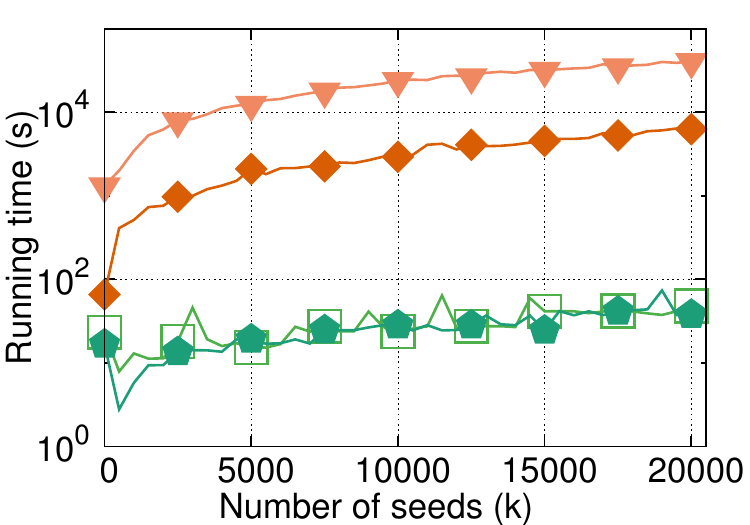}
	}
	\vspace{-0.1in}
	\caption{Running time under LT model}
	\label{fig:time_lt}
	\vspace{-0.2in}
\end{figure*}

\begin{figure*}[!ht]
	\subfloat[NetHEPT]{
		\includegraphics[width=0.24\linewidth]{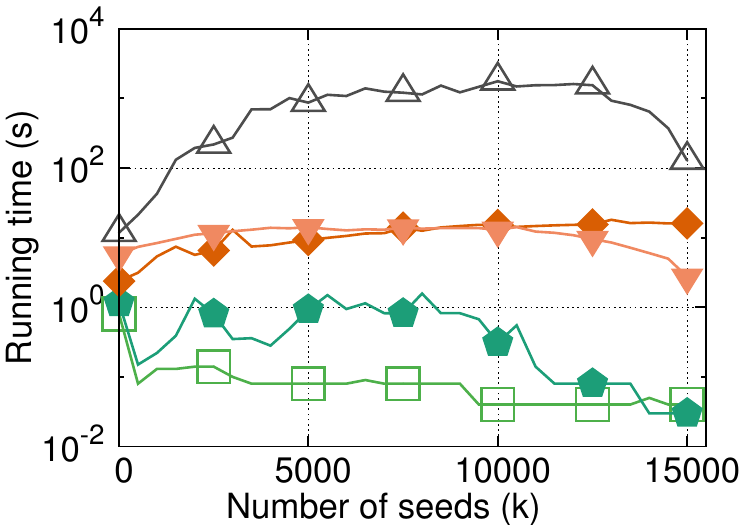}
	}
	\subfloat[NetPHY]{
		\includegraphics[width=0.24\linewidth]{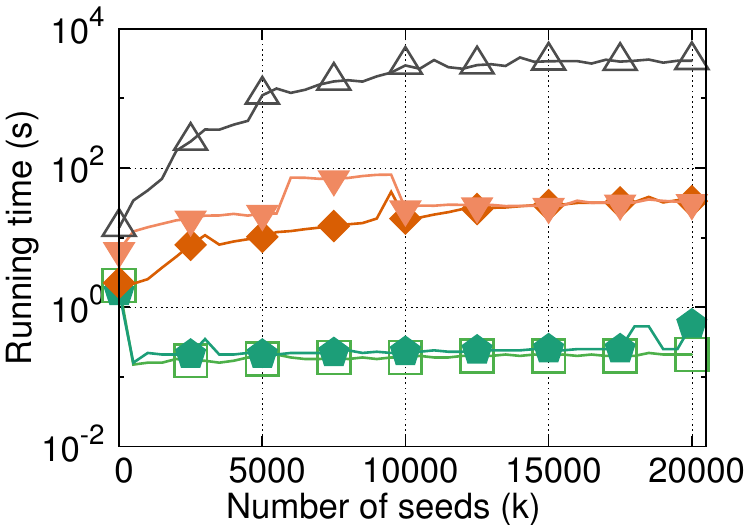}
	}
	\subfloat[DBLP]{
		\includegraphics[width=0.24\linewidth]{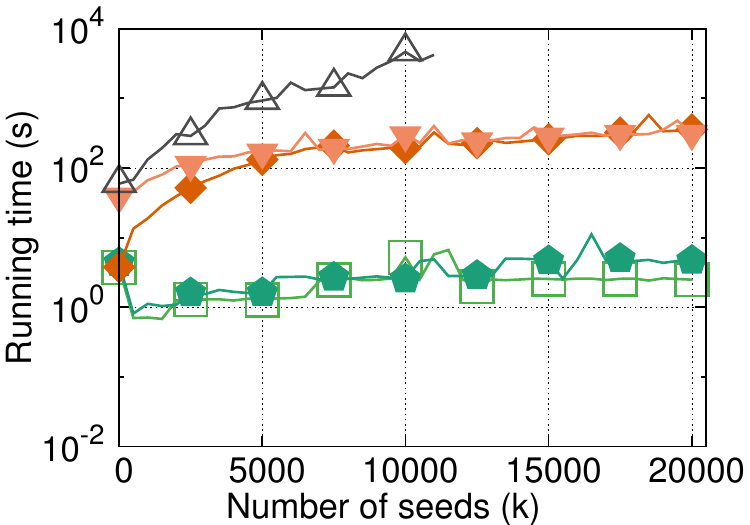}
	}
	\subfloat[Twitter]{
		\includegraphics[width=0.24\linewidth]{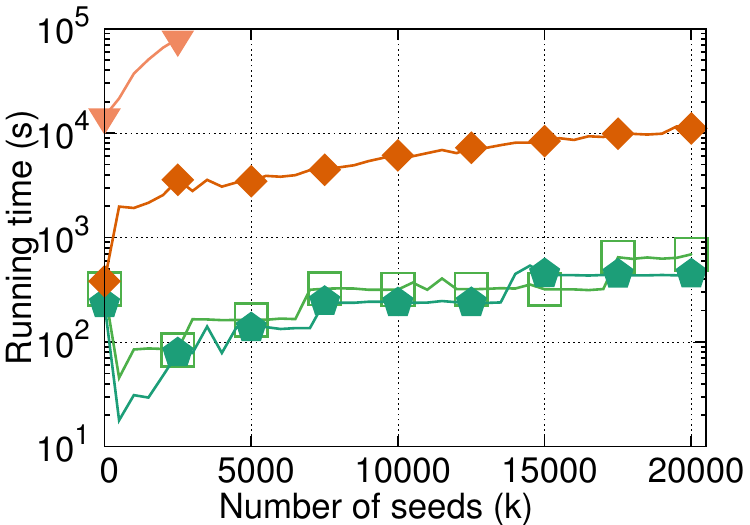}
	}
	\vspace{-0.1in}
	\caption{Running time under IC model}
	\label{fig:time_ic}
	\vspace{-0.2in}
\end{figure*}

\begin{figure*}[!t]
	\subfloat[NetHEPT]{
		\includegraphics[width=0.24\linewidth]{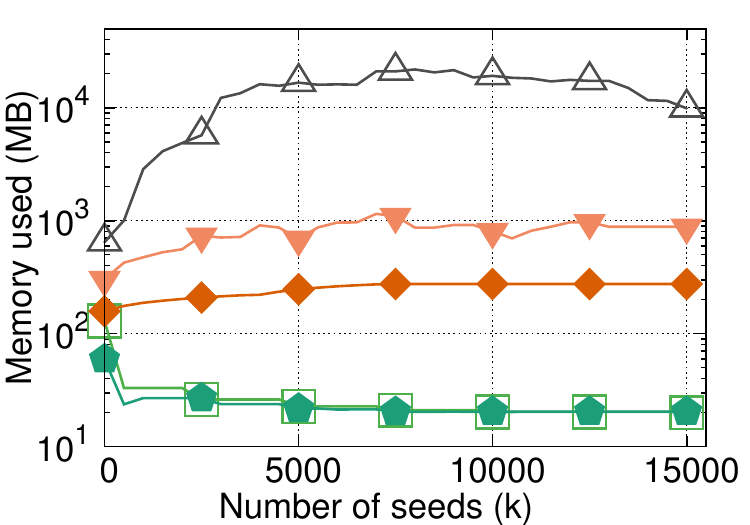}
	}
	\subfloat[NetPHY]{
		\includegraphics[width=0.24\linewidth]{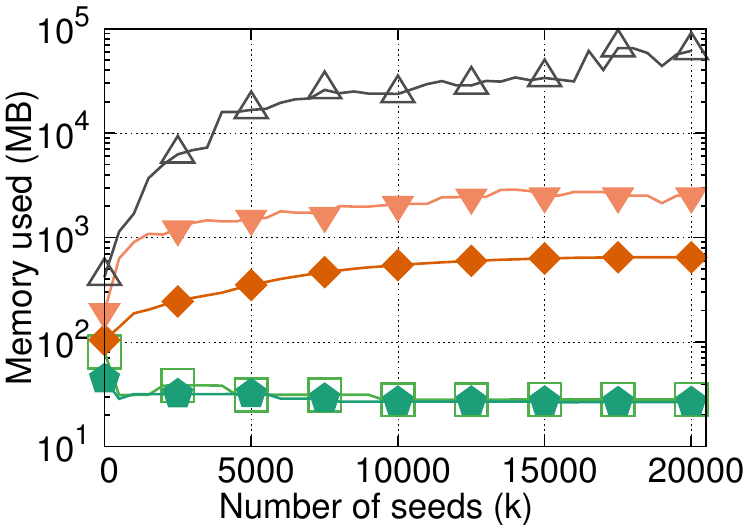}
	}
	\subfloat[DBLP]{
		\includegraphics[width=0.24\linewidth]{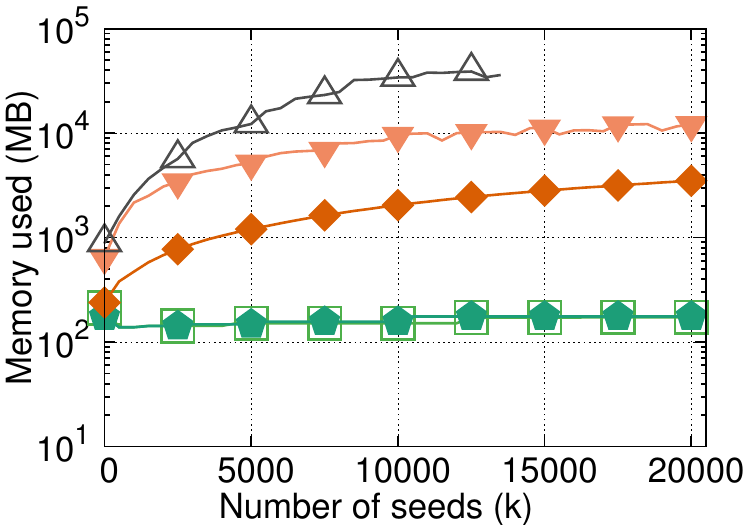}
	}
	\subfloat[Twitter]{
		\includegraphics[width=0.24\linewidth]{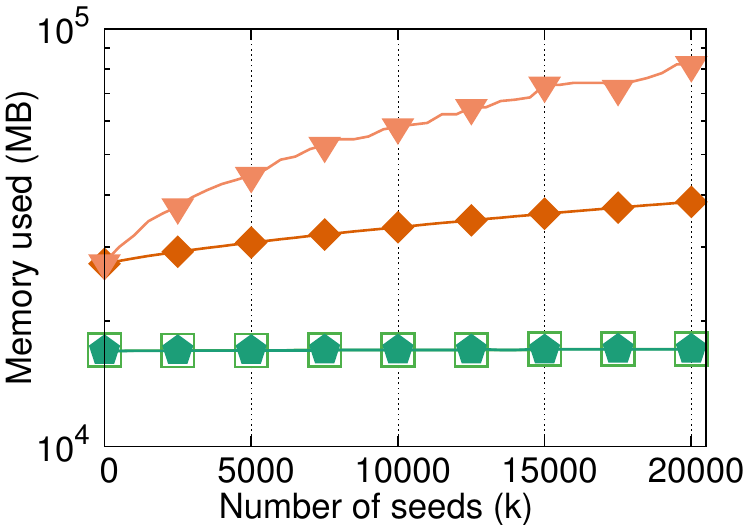}
	}
	\vspace{-0.1in}
	\caption{Memory usage under LT model}
	\label{fig:mem_lt}
	\vspace{-0.2in}
\end{figure*}

\begin{figure*}[!t]
	\subfloat[NetHEPT]{
		\includegraphics[width=0.24\linewidth]{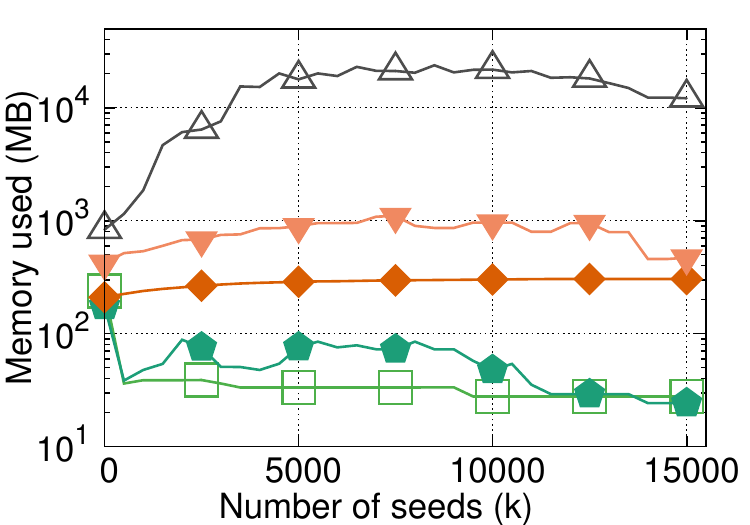}
	}
	\subfloat[NetPHY]{
		\includegraphics[width=0.24\linewidth]{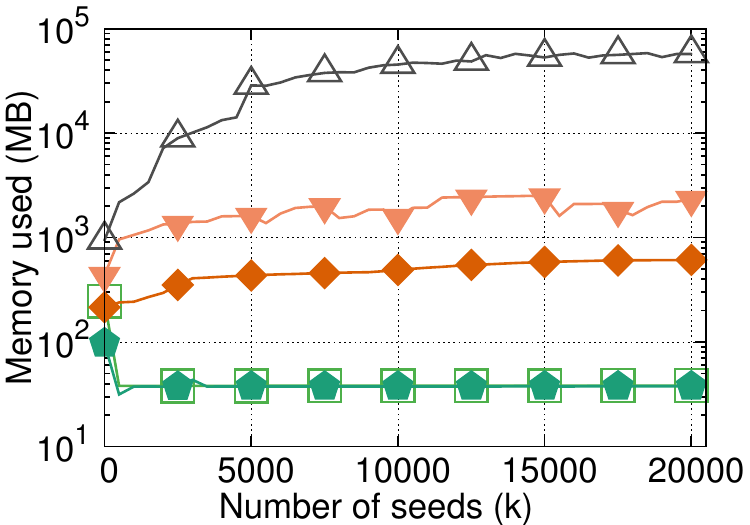}
	}
	\subfloat[DBLP]{
		\includegraphics[width=0.24\linewidth]{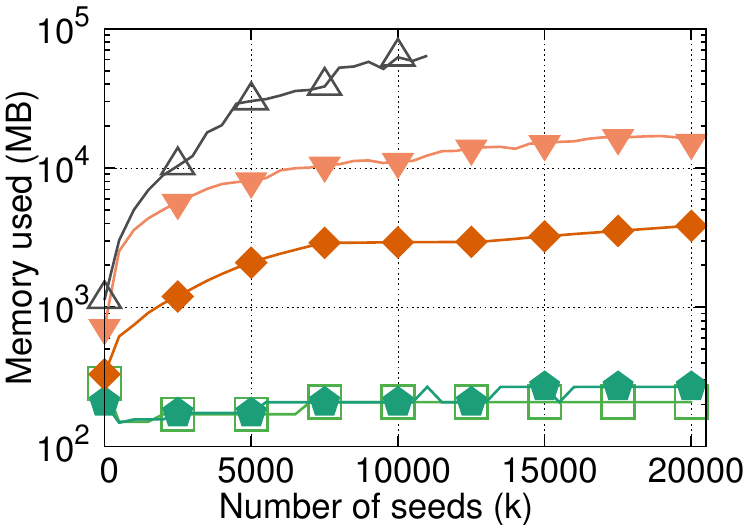}
	}
	\subfloat[Twitter]{
		\includegraphics[width=0.24\linewidth]{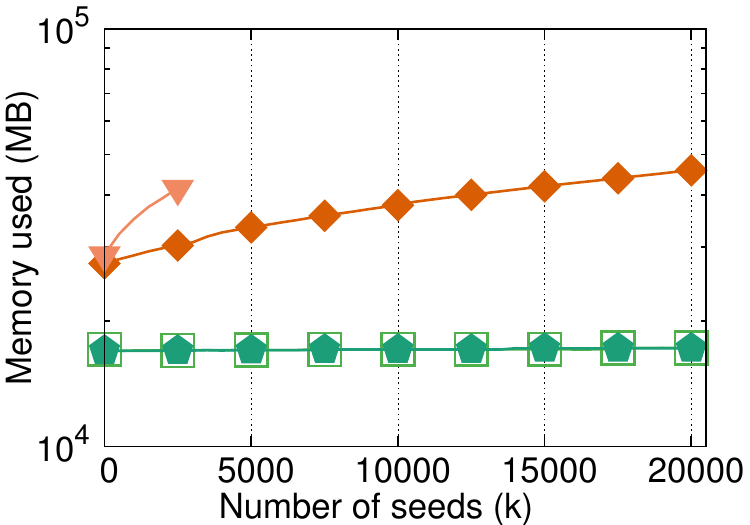}
	}
	\vspace{-0.1in}
	\caption{Memory usage under IC model}
	\label{fig:mem_ic}
\end{figure*}
\subsection{Experiments with IM problem}

To show the superior performance of the proposed algorithms on \IM{} task, we ran the first set of experiments on four real-world networks, i.e., NetHEPT, NetPHY, DBLP, Twitter. We also test on a wide spectrum of the value of $k$, typically, from 1 to 20000, except on NetHEPT network since it has only 15233 nodes. The solution quality, running time, memory usage are reported sequentially in the following. We also present the actual number of RR sets generated by \PIMA{}, \DPIMA{} and IMM when testing on four other datasets, i.e., Enron, Epinions, Orkut and Friendster.
\subsubsection{Solution Quality}
We first compare the quality of the solution returned by all the algorithms on LT and IC models. The results are presented in Fig.~\ref{fig:qual_inf_lt} and Fig.~\ref{fig:qual_inf_ic}, respectively. The CELF++ algorithm is only able to run on NetHEPT due to time limit. From those figures, all the methods return comparable seed set quality with no significant difference. The results directly give us a better viewpoint on the basic network property that a small fraction of nodes can influence a very large portion of the networks. Most of the previous researches only find up to 50 seed nodes and provide a limited view of this phenomenon. Here, we see that after around 2000 nodes have been selected, the influence gains of selecting more seeds become very slim.

\setlength\tabcolsep{1pt}
\begin{table*}[!t]\centering
    \begin{tabular}{l|rrr|rrr|rrr|rrr|rrr|rrr}
		\toprule
		\multirow{3}{*}{Data}& \multicolumn{9}{c|}{\textbf{running time}} & \multicolumn{9}{c}{\textbf{number of RR sets}} \\
		\cline{2-10} \cline{11-19}
		& \multicolumn{3}{c|}{$k=1$} & \multicolumn{3}{c|}{$k=500$} & \multicolumn{3}{c|}{$k=1000$} & \multicolumn{3}{c|}{$k=1$} & \multicolumn{3}{c|}{$k=500$} & \multicolumn{3}{c}{$k=1000$}\\
		\cline{2-4} \cline{5-7} \cline{8-10} \cline{11-13} \cline{14-16} \cline{17-19}
		& \DPIMA & \PIMA & \multicolumn{1}{r|}{IMM} & \DPIMA & \PIMA & \multicolumn{1}{r|}{IMM} & \DPIMA & \PIMA & \multicolumn{1}{r|}{IMM} & \DPIMA & \PIMA & \multicolumn{1}{r|}{IMM} & \DPIMA & \PIMA & \multicolumn{1}{r|}{IMM} & \DPIMA & \PIMA & IMM\\
		\midrule
		Enron & \textbf{.5 s} & .6 s & \multicolumn{1}{r|}{.7 s}& \textbf{.1 s} & \textbf{.1 s} & \multicolumn{1}{r|}{3.1 s}&\textbf{.1} s& \textbf{.1 s}  & \multicolumn{1}{r|}{6.9 s} & \textbf{96 K} & 272 K & \multicolumn{1}{r|}{280 K} & \textbf{24 K}& 42 K& \multicolumn{1}{r|}{580 K} & \textbf{24 K} & 61 K & 910 K\\
		Epin.& \textbf{.6 s}& .7 s & \multicolumn{1}{r|}{.8 s} & \textbf{.2 s}& \textbf{.2 s}& 4.4 s& \textbf{.2 s} & .3 s & 12.1 s& \textbf{205 K}  & 570 K& 400 K& \textbf{51 K}& 97 K & 1.2 M& \textbf{51 K} & 131 K& 1.9 M\\
		Orkut & \textbf{86.2 s} & 108.2 s& \multicolumn{1}{r|}{179.9 s}& \textbf{11.8 s} & 12.1 s& 317.8 s& \textbf{23.8 s}& 25.8 s& 548.9 s& \textbf{512 K} & 1.5 M & 1.2 M& \textbf{64 K} & 177 K & 2.1 M& \textbf{128 K} & 230 K& 3.3 M \\
		Frien. & \textbf{4.1 h}&\  4.7 h  & 8.1 h & \textbf{.27 h}& .48 h & n/a & \textbf{.26 h} & .48 h & n/a & \textbf{77 M} & 161 M & 175 M &\textbf{4.8 M}&\ 17 M & n/a & \textbf{4.8 M} &\ 15 M & n/a \\

		\bottomrule

	\end{tabular}
	\vspace{-0.1in}
	\caption{Performance of \PIMA{}, \DPIMA{} and IMM on various datasets under LT model.}
	\label{tab:res}
\end{table*}

\subsubsection{Running time} We next examine the performance in terms of running time of the tested algorithms. The results are shown in Fig.~\ref{fig:time_lt} and Fig.~\ref{fig:time_ic}. Both \PIMA{} and \DPIMA{} significantly outperform the other competitors by a huge margin. Comparing to IMM, the best known algorithm, \PIMA{} and \DPIMA{} run up to several orders of magnitudes faster. TIM+ and IMM show similar running time since they operate on the same philosophy of estimating optimal influence first and then calculating the necessary samples to guarantee the approximation for all possible seed sets. However, each of the two steps displays its own weaknesses. In contrast, \PIMA{} and \DPIMA{} follows the Stop-and-Stare mechanism to thoroughly address those weaknesses and thus exhibit remarkable improvements. In particular, the speedup factor of \DPIMA{} to IMM can go up to 1200x in the case of NetHEPT network on the LT model. On most of other cases, the factor stabilizes at several hundred times.

Comparing between \PIMA{} and \DPIMA{}, since \DPIMA{} possesses the type-2 minimum threshold compared to the weaker type-1 threshold of \PIMA{} with the same precision settings $\epsilon,\delta$, \DPIMA{} performs at least as good as \PIMA{} and outperforms in many cases in which the fixed setting of \PIMA{} falls out of the effective ranges for that network and value $k$. This problem is resolved in \DPIMA{} thanks to the dynamic error computation at every iteration.

\subsubsection{Memory Usage and Number of RR sets} 
This experiment is divided into two parts: 1) we report the memory usage in the previous experiments and 2) since the gain in influence peaks at the selection of 1 to 1000 nodes, we carry new experiments on four other datasets, i.e., Enron, Epinion, Orkut and Friendster, with $k \in \{1, 500, 1000\}$ to show the view across datasets of \PIMA{}, \DPIMA{} and IMM.

\textit{Memory Usage.} The results on memory usage of all the algorithms are shown in Fig.~\ref{fig:mem_lt} and Fig.~\ref{fig:mem_ic}. We can see that there is a strong correlation between running time and memory usage. It is not a surprise that \PIMA{} and \DPIMA{} require much less memory, up to orders of magnitude, than the other methods since the complexity is represented by the number of RR sets and these methods achieve type-1 and type-2 minimum thresholds of RR sets.

\textit{Across datasets view.} We ran \PIMA{}, \DPIMA{} and IMM on four other datasets, i.e., Enron, Epinions, Orkut and Friendster, with $k \in \{1, 500, 1000\}$ under LT model. The results are presented in Table \ref{tab:res}. In terms of running time, the table reflects our previous results that \PIMA{} and \DPIMA{} largely outperform IMM, up to several orders of magnitudes. The same pattern happens in terms of the number of RR sets generated. As shown, even in the most extreme cases of selecting a single node, \PIMA{} and \DPIMA{} require several times fewer RR sets than IMM.

We note that, in the most challenging case of Friendster network with over 3.6 billion edges, IMM uses 172 GB of main memory while \DPIMA{} and \PIMA{} require much lower memory resource of only 69 and 72 GB respectively.
\subsection{Experiments with TVM problem}
In this experiments, we will modify our Stop-and-Stare algorithms to work on Targeted Viral Marketing (TVM) problem and compare with the best existing method, i.e., KB-TIM in \cite{Li15} to show the drastic improvements when applying our methods. In short, we will describe how we select the targeted groups from actual tweet/retweet datasets of Twitter and how to modify \DPIMA{} and \PIMA{} for TVM problem. Then, we will report the experimental results.
\subsubsection{TVM problem and methods}
Targeted Viral Marketing (TVM) is a central problem in economics in which, instead of maximizing the influence over all the nodes in a network as in \IM{}, it targets a specific group whose users are relevant to a certain topic and aims at optimizing the influence to that group only. Each node in the targeted group is associated with a weight which indicates the relevance of that user to the topic. The best current method for solving TVM is proposed in \cite{Li15} in which the authors introduce weighted \RIS{} sampling (called WRIS) and integrate it into TIM+ method \cite{Tang14} to derive an approximation algorithm, termed KB-TIM. WRIS only differs from the original \RIS{} at the point of selecting the sampling root. More specifically, WRIS selects the root node proportional to the node weights instead of uniform selection as in \RIS{}.

In the same way, we incorporate WRIS into \DPIMA{} and \PIMA{} for solving TVM problem. By combining the analysis of WRIS in \cite{Li15} and our previous proofs, it follows that the modified \DPIMA{} and \PIMA{} preserve the $(1-1/e-\epsilon)$-approximation property as in \IM{} problem.
\subsubsection{Extracting the targeted groups}
\setlength\tabcolsep{5pt}
\begin{table}[!htb]
	\caption{Topics, related keywords}
	\label{tab:topic}
	\centering
	\begin{tabular}{c|p{5cm}|r}
		\toprule
		\emph{Topic} & \emph{Keywords}& \emph \#Users\\
		\midrule
		1& \textbf{bill clinton, iran, north korea, president obama, obama} & 997,034\\
		\hline
		2& \textbf{senator ted kenedy, oprah, kayne west, marvel, jackass} & 507,465\\
		\bottomrule
	\end{tabular}
\end{table}
We use tweet/retweet dataset to extract the users' interests on two political topics as described in \cite{Kwak10}. We choose two groups of most popular keywords as listed in Table~\ref{tab:topic}, and mine from the tweet data who posted tweets/reweets containing at least one of those keywords in each group and how many times. We consider those users to be the targeted groups in TVM experiments with the relevance/interest of each user on the topic proportional to the frequency of having those keywords in their tweets.
\subsubsection{Experimental results}
\begin{figure}[!h]
	\centering
	\subfloat[Topic 1]{
		\includegraphics[width=0.48\linewidth]{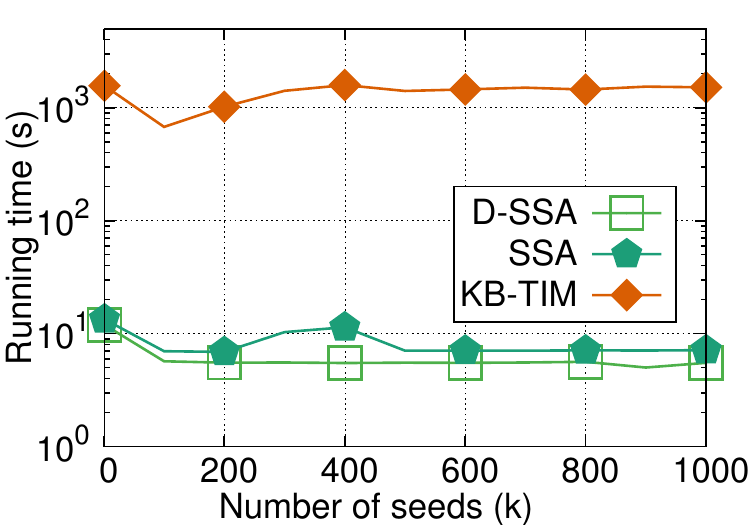}
		\label{fig:tvm_time_1}
	}
	\subfloat[Topic 2]{
		\includegraphics[width=0.48\linewidth]{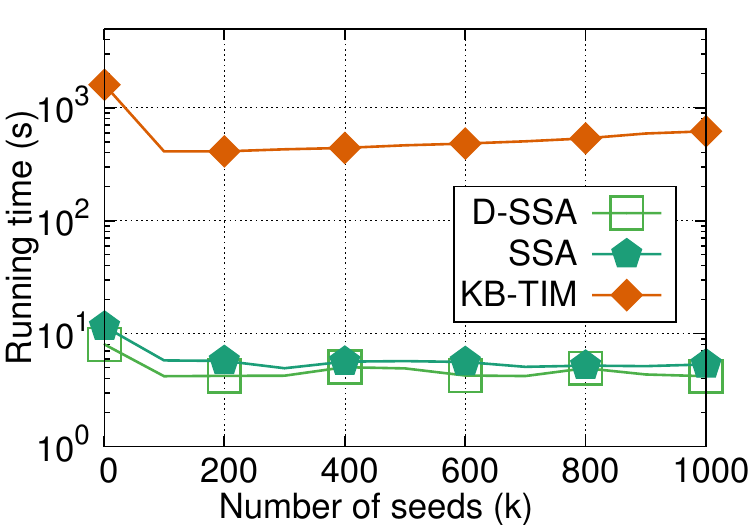}
		\label{fig:tvm_time_2}
	}
	\caption{Running time on Twitter network}
	\label{fig:tvm_time}
\end{figure}

We run \PIMA{}, \DPIMA{} and KB-TIM on Twitter network under LT model with the targeted groups extracted from tweet dataset as described previously. Since all the algorithms have the same guarantee on the returned solution, we only measure the performance of these methods in terms of running time and the results are depicted in Fig.~\ref{fig:tvm_time}. In both cases, \DPIMA{} and \PIMA{} consistently witness at least two order of magnitude improvements (up to 500 times) in running time compared to KB-TIM. \DPIMA{} is also consistently faster than \PIMA{} due to the more optimal type-2 threshold.

\section{Conclusion}
\label{sec:con}
In this paper, we make several significant contributions in solving the fundamental influence maximization (\IM{}) problem. We provide the unified \RIS{} framework which generalizes the best existing technique of using \RIS{} sampling to find an $(1-1/e-\epsilon)$-approximate solution in billion-scale networks. We introduce the \RIS{} threshold that all the algorithms following the framework need to satisfy and two minimum thresholds, i.e., type-1 and type-2. Interestingly, we are able to develop two novel algorithms, \PIMA{} and \DPIMA{}, which are the first methods meeting the two minimum thresholds. Since \IM{} plays a central roles in a wide range of practical applications, e.g., viral marketing, controlling diseases, virus/worms, detecting contamination and so on, the developments of \PIMA{} and \DPIMA{} will immediately result in a burst in performance and allow their applications to work in billion-scale domains.
\section{Acknowledgments}
The work of My T. Thai is partially supported by NSF CCF 1422116.
\bibliographystyle{ieeetr} 

\appendix

%
%
%
%
	\section*{Tightness of Chernoff's bounds}
	In the following proofs, we use an intermediate results on the optimality of Chernoff-like in Lem.~\ref{lem:chernoff}. According to Lemma~4 in \cite{Klein15}, we have the following results regarding the tightness of the Chernoff-like bounds in the above lemma.
	\begin{Lemma}[\cite{Klein15}]
		\label{lem:tightness}
		Let $X_1, X_2, \dots,..., X_T$ be i.i.d random variables taking values 0 or 1, and $\Pr[X_i = 1] = \mu \leq 1/2$. Denote by $\hat \mu = \frac{1}{T}\sum_{i=1}^T X_i$ the average of the random variables. For every $\epsilon \in (0, 1/2]$, if $\epsilon^2 \mu T \geq 3$, the following hold:
		\begin{align}
			&\Pr[\hat \mu \leq (1-\epsilon)\mu] \geq \exp{(-9 \epsilon^2\mu T)} \text{ and }\\
			&\Pr[\hat \mu \geq (1+\epsilon)\mu] \geq \exp{(-9 \epsilon^2\mu T)}.
		\end{align}
	\end{Lemma}
Note that the condition $\epsilon \in (0, 1/2]$ can be relaxed into $\epsilon \in (0, c]$ for any $c < 1$ if we assume sufficiently small $\delta$. 

\begin{corollary}[Tightness of Chernoff's bound ]
		\label{col:optimal0}
Let $X_1, X_2, \dots $ be i.i.d random variables taking values $0$ or $1$, and $\Pr[X_i = 1] = \mu \in(0, 1/2]$. For $\epsilon \in (0, 1/2]$, $\delta < 1/e$ and $T>0$, the following hold
\begin{itemize}
	\item If $\Pr \left[ \frac{1}{T}\sum_{i=1}^{T}X_i  < (1-\epsilon)\mu \right] \leq \delta$, then 	$T = \Omega( \Upsilon(\epsilon, \delta) \frac{1}{\mu} )$.
	\item If $\Pr \left[ \frac{1}{T}\sum_{i=1}^{T}X_i  > (1+\epsilon)\mu \right] \leq \delta$, then 	$T = \Omega( \Upsilon(\epsilon, \delta) \frac{1}{\mu} )$.
\end{itemize}
\end{corollary}
\begin{proof}
	If $T < \frac{1}{9} \frac{1}{\epsilon^2} \ln \frac{1}{n}$,
	 then by Lem.~\ref{lem:tightness}, $\Pr[\hat \mu \leq (1-\epsilon)\mu] \geq  \exp(-9 \epsilon^2\mu T) = \delta$ (contradiction). Thus, $T\geq  \frac{1}{9} \frac{1}{\epsilon^2} \ln \frac{1}{n}=\Omega(\Upsilon(\epsilon, \delta))$.
	 
	  Similarly, if $T < \frac{1}{9} \frac{1}{\epsilon^2} \ln \frac{1}{n}$, then 
$\Pr[\hat \mu \geq (1+\epsilon)\mu] \geq \exp{(-9 \epsilon^2\mu T)} = \delta$ (contradiction). Thus,
\[\Pr \left[ \frac{1}{T}\sum_{i=1}^{T}X_i  > (1+\epsilon)\mu \right] \leq \delta\] implies $T\geq \frac{1}{9} \frac{1}{\epsilon^2} \ln \frac{1}{n} =\Omega(\Upsilon(\epsilon, \delta))$.
\end{proof}	
	
	The lower bounds also hold for the case when $X_1, \ldots, X_T$ are weakly dependent (martingales) as the random variables in Lem.~\ref{lem:chernoff}. 

	\subsection*{Proof of Theorem~\ref{theo:type1}}
		Apply the union bound. The following two inequalities from Eqs.~\ref{eq:sk} and~\ref{eq:sk*} hold together with probability at least $1- (\delta_a+\delta_b)$.
		\begin{align}
		\label{eq:epsa}\hat \I(\hat S_k) \leq (1 + \epsilon_a) \I(\hat S_k) \\
		\label{eq:epsb}\I(\hat S^*_k) \geq  (1-\epsilon_b)\I(S^*_{k}).
		\end{align}
		\textit{Assume that the above two inequalities hold}. We show, by contradiction, that $\I(\hat S_k) \geq (1-1/e-\epsilon)\OPT_k$, where $\epsilon = (1-\frac{1}{e})\frac{\epsilon_a + \epsilon_b}{1+\epsilon_a}$. Assume  the opposite, i.e., 
		\begin{align}
		\label{eq:contrad1}
		\I(\hat S_k) < (1-1/e-\epsilon)\OPT_k. 
		\end{align}
		
		Since the greedy algorithm used in \textsf{Max-Coverage} algorithm returns a $(1-1/e)$ approximation  \cite{Nemhauser81}, the greedy solution $\hat S_k$ satisfies $\Cov_{\mathcal R}(\hat S_k) \geq (1-1/e) \Cov_{\mathcal R}(S^*_k)$. It follows that
		\[
		\hat \I(\hat S_k) \geq (1 - 1/e)\hat \I(S^*_k).
		\]
		Extend (\ref{eq:epsa}) and use the assumption (\ref{eq:contrad1}).
		\begin{align}
		\nonumber  \I(\hat S_k) &\geq \hat \I(\hat S_k) - \epsilon_a 	 \I(\hat S_k) \geq (1 - 1/e)\hat \I(S^*_{k}) - \epsilon_a 	 \I(\hat S_k)\\
		&\geq (1 - 1/e)\hat \I(S^*_{k}) - \epsilon_a (1-1/e-\epsilon)	 \OPT_k
		\end{align}
		Apply Eq. (\ref{eq:epsb}), we yield
		\begin{align*}
		\nonumber  \I(\hat S_k) 	&\geq 		
		(1 - 1/e)(1-\epsilon_b)\I(S^*_{k}) - \epsilon_a (1-1/e-\epsilon)	 \OPT_k\\
		&= (1 - 1/e - (1-1/e -\epsilon)\epsilon_a+(1-1/e)\epsilon_b)\OPT_k \\
		&= (1 - 1/e - \epsilon)\OPT_k \text{  \textit{(contradiction)}}
		\end{align*}
		where $\epsilon =(1-1/e -\epsilon)\epsilon_a+(1-1/e)\epsilon_b)$, or equivalently, $\epsilon = (1-\frac{1}{e})\frac{\epsilon_a + \epsilon_b}{1+\epsilon_a}$.
		
		Thus, $\Pr[ \I(\hat S_k) \geq  (1 - 1/e - \epsilon)OPT_k ] \geq  1 -(\delta_a+\delta_b)$.

	\subsection*{Proof of Lemma~\ref{cor:check}}
	
	We follow the proof of the \textsf{Stopping Rule Theorem} in \cite{Dagum00}.
	
	Since $\I_c (\hat S_k) = n \Lambda_2/T$, it suffices to show that
	\begin{align}
		\Pr[T \leq \frac{n\Lambda_2}{(1+\epsilon_2)\I(\hat S_k)}] \leq \frac{\delta_2}{3\log_2n},
	\end{align}
	where $T \leq T_{max}$ is the number of RR sets generated.

	Let $L = \floor{\frac{n \Lambda_2}{(1+\epsilon_2) \I(\hat S_k)}}$. From the definition of $\Lambda_2$, we obtain that,
	\begin{align}
		L & = \floor{\frac{n (1+(1+\epsilon_2)(2+\frac{2}{3}\epsilon_2) \ln(\frac{1}{\delta'_2})\frac{1}{\epsilon_2^2})}{(1+\epsilon_2) \I(\hat S_k)}} \\
		& \geq (2+\frac{2}{3}\epsilon_2) \ln(\frac{1}{\delta'_2}) \frac{n}{\I(\hat S_k) \epsilon_2^2}.
	\end{align}
	
	Since $T$ is an integer, $T \leq \frac{n \Lambda_2}{(1+\epsilon_2) \I(\hat S_k)}$ if and only if $T \leq L$. But $T \leq L$ if and only if $\Cov_L = \sum_{j=1}^{L} Z_j \geq \Lambda_2$. Thus,
	\begin{align}
		\Pr[T \leq \frac{n \Lambda_2}{(1+\epsilon_2) \I(\hat S_k)}] & = \Pr[T \leq L] = \Pr[\Cov_L \geq \Lambda_2] \\
		& = \Pr[\Cov_L n / L \geq \Lambda_2 n / L] \\
		& \leq \Pr[\hat \mu_L \geq (1+\epsilon_2) \mu].
	\end{align}
	Apply the Chernoff's bound in Lem.~\ref{lem:chernoff} on the last probability and note that $L \geq (2+\frac{2}{3}\epsilon_2) \ln(\frac{1}{\delta'_2}) \frac{n}{\I(\hat S_k) \epsilon_2^2}$, we achieve the following bound,
	\begin{align}
		\Pr[\hat \mu_L \geq (1+\epsilon_2) \mu] \leq \delta'_2 = \frac{\delta_2}{3 \log_2 n}.
	\end{align} 
	
	Thus, we have,
	\begin{align}
		\Pr[\I_c(\hat S_k) \geq (1+\epsilon_2) \I(\hat S_k)] \leq \frac{\delta_2}{3 \log_2 n},
	\end{align}
	which completes the proof of Lem.~\ref{cor:check}.
	
\balance
	
	\subsection*{Proof of Lemma~\ref{lem:com2}}
%
		Note that there are $|\R|$ RR sets to estimate the influence of the optimal solution $S^*_k$. We use Chernoff-Hoeffding's inequality (Lem.~\ref{lem:chernoff}) on the optimal solution, $S^*_k$, with random variable $Z = \min\{1,|R_j \cap S^*_k|\}$ and $\mu_Z = \OPT_k/n$ to obtain
		\begin{align}
		\Pr [\hat \I(S^*_k)\leq (1 - \epsilon^{(i)}_3)\OPT_k] \leq e^{-\frac{|\R| \OPT_k(\epsilon^{(i)}_3)^2}{2 n}} \leq \delta/(3i_{max}),
		\end{align}
		which completes the proof of Lem. \ref{lem:com2}.
		
\subsection*{Proof of Theorem~\ref{theo:approx}}
	Assume that none of the bad events in Lemmas \ref{lem:caps}, \ref{lem:com1}, and \ref{lem:com2} happens. By union bound, this assumption holds with probability at least
	\[ 1 - (\delta/3 +  \delta/(3i_{max}) \times 3i_{max} + \delta/(3i_{max}) \times 3i_{max})  = 1-\delta.
	\]	
	We will show that $\I(\hat S_k) \geq (1 - 1/e - \epsilon) \OPT_k$.
	
	If \PIMA{} terminates with $|\R|\geq N_{max}$, since the bad event  $\left[|R| \geq N_{max} \text{ and }\I( \hat S_k) < (1-1/e - \epsilon)\right]$ (Lem. \ref{lem:caps}) does not happen, we have $\I(\hat S_k) \geq (1 - 1/e - \epsilon) \OPT_k$.
	
	Otherwise, \PIMA{} will stop due to the two stopping conditions (C1), Line 8 Alg.~\ref{alg:main}, and (C2), Line 11 Alg.~\ref{alg:main}.
	
	
	\textit{Proving $\epsilon_3^{(t)} \leq \epsilon_3$}. Since the bad event in Lem.~\ref{lem:com1} does not happen, we have
	\[
	\I_c(\hat S_k) \leq (1+\epsilon_2)\I(\hat S_k).
	\]
	Thus,
	\begin{align}
	\label{eq:e12}
	\hat \I(\hat S_k) = (1+\epsilon_1)\I_c(\hat S_k) \leq (1+\epsilon_1)(1+\epsilon_2)\I(\hat S_k).
	\end{align}
	From the stopping condition (C1) $\Cov_{\R}(\hat S_k) \geq \Lambda_1$, we have
	\begin{align}
	\nonumber		\hat \I(\hat S_k) &= \frac{\Cov_{\R}(\hat S_k) n}{|\R|} 
	\geq \frac{\Lambda_1 n}{|\R|} = \frac{(1 + \epsilon_1)(1 + \epsilon_2) \Upsilon(\epsilon_3,\frac{\delta}{3i_{max}}) n}{|\R|}\\
	\label{eq:ptR}\Rightarrow |\R| &\geq 	\frac{\Lambda_1 n}{\hat \I(\hat S_k)} = \frac{(1 + \epsilon_1)(1 + \epsilon_2) \Upsilon(\epsilon_3,\frac{\delta}{3i_{max}}) n}{\hat \I(\hat S_k)}	
	\end{align}		
	Combine with Eq. (\ref{eq:e12}), we obtain
	\begin{align*}
	|\R| \geq 	 \frac{(1 + \epsilon_1)(1 + \epsilon_2) \Upsilon(\epsilon_3,\frac{\delta}{3i_{max}}) n}{(1+\epsilon_1)(1+\epsilon_2)\I(\hat S_k)}	\\
	= \frac{ \Upsilon(\epsilon_3,\frac{\delta}{3i_{max}}) n}{\I(\hat S_k)} \geq \frac{ \Upsilon(\epsilon_3,\frac{\delta}{3i_{max}}) n}{\OPT_k}.
	\end{align*}
	Substitute the above into the definition of $\epsilon_3^{(t)}$. We have
	\begin{align}
	\epsilon_3^{(t)} = \sqrt{\frac{2n \ln \frac{3i_{max}}{\delta}}{|\R| \OPT_k}} \leq \sqrt{\frac{2n \ln \frac{3i_{max}}{\delta}}{\frac{\Upsilon(\epsilon_3,\frac{\delta}{3i_{max}}) n}{\OPT_k} \OPT_k}} \leq \epsilon_3.
	\end{align}
	
	\textit{Proving the approximation ratio}. 	
	Combine the above with the assumption that the bad event in the Lem.~\ref{lem:com2} does not happen, we have
	\begin{align}
	\label{eq:eq23}
	\hat \I(S^*_k) \geq (1-\epsilon_3^{(t)}) \OPT_k \geq (1-\epsilon_3) \OPT_k.
	\end{align}
	Let $\epsilon_a = \epsilon_1 + \epsilon_2 + \epsilon_1 \epsilon_2$. We can rewrite Eq. (\ref{eq:e12}) into  \[\hat \I(\hat S_k) \leq (1 + \epsilon_a) \I(\hat S_k).\]
	Follow the same contradiction proof in the Theorem~\ref{theo:type1} with $\epsilon_a$ and $\epsilon_b = \epsilon_3$, we have $\I(\hat S_k) \geq (1-1/e - \epsilon) \OPT_k$, where $\epsilon = (1-\frac{1}{e})\frac{\epsilon_a+\epsilon_b}{1+\epsilon_a}=(1-\frac{1}{e})\frac{\epsilon_1+\epsilon_2+\epsilon_1\epsilon_2+\epsilon_3}{(1+\epsilon_1)(1+\epsilon_2)}$.
	
	Therefore, $\Pr[\I(\hat S_k) \geq (1-1/e - \epsilon) \OPT_k] \geq 1 - \delta$.
	
\subsection*{Proof of Lemma~\ref{lem:tssa}}
	Since $|\R| \geq T_1$ implies
	\[
	\Pr[ \hat \I_{\R}(S^*_k)  \geq (1-\epsilon_b)OPT_k] \geq 1 - \delta_b.
	\]
	From the assumption that  $1/\delta =\Omega(\ln n)$ and the fact that $i_{max} \leq 2\log_2 n$ and $\delta_b \leq \delta$ , we have
	\begin{align}
	\nonumber&\Upsilon(\epsilon_0, \frac{\delta}{3 i_{max}})= (2+2/3 \epsilon_0) \frac{1}{\epsilon_0^2} \ln \frac{3 i_{max}}{\delta}\\
	\label{eq:tssac}&\leq  3 \frac{\epsilon_b^2}{\epsilon_0^2} \frac{1}{\epsilon_b^2} (\ln 1/\delta + \ln 3 i_{max} )
	= O( \Upsilon(\epsilon_b, \delta_b) )
	\end{align}
	The values of $\epsilon_2, \epsilon_3$ specified later at the end of Theorem~\ref{theo:efficiencyssa} will guarantee that $\frac{\epsilon_b^2}{\epsilon_0^2}$ is also a constant that depends only on $\epsilon_b$.
	
	Apply Corollary~\ref{col:optimal0}, we have $T_1 = \Omega( \Upsilon(\epsilon_b, \delta_b) \frac{n}{\OPT_k})$. Thus,
	\begin{align}
	\nonumber T_{\PIMA} &= \max\{ T_1, \alpha \Upsilon(\epsilon_0, \frac{\delta}{3i_{max}}) \frac{n}{\OPT_k}\} \\
	&= O(\Upsilon(\epsilon_b, \delta_b))\frac{n}{\OPT_k} = O( T_1)
	\end{align}	
	This yields the proof.
	
\subsection*{Proof of Theorem~\ref{theo:efficiencyssa}}
	\PIMA{} stops when either $|\R| \geq N_{max}$ or all the following stopping conditions hold simultaneously.
	\begin{itemize}
		\item $\Cov_{\R}(\hat S_k) \geq \Lambda_1=\Theta(\Upsilon(\epsilon_3,\frac{\delta}{3i_{max}}))$ (Condition C1)
		\item \textsf{Estimate-Inf}($G, \hat S_k, \epsilon_2, \delta'_2$, $T_{max}$) returns an estimation $\I_c(\hat S_k)$ but not $-1$. (Line 10, Alg.~\ref{alg:main}).
		\item $\hat \I(\hat S_k) \leq (1+\epsilon_1) \I_c(\hat S_k)$ (Condition C2).
	\end{itemize}

	Assume $|\R| \geq 	T_{\PIMA} = O(T_1)$ (Lem.~\ref{lem:tssa}). If $T_{\PIMA} \geq N_{max}$, then \PIMA{} will stop within $ O(T_1)$ samples. Otherwise,  $|\R| \geq T_{\PIMA}$ at some iteration $i \leq i_{max}$.

	Assume that none of the bad events in Lemmas \ref{lem:com1}, \ref{lem:com2}, \ref{lem:eit}, and Eqs. (\ref{eq:eq1}), and Eq. (\ref{eq:eq2})  happen. By union bound, the assumption holds with a probability at least
	\[ 1- (\frac{\delta}{3i_{max}}  3i_{max} + \frac{\delta}{3i_{max}}  3i_{max}+ \frac{\delta}{3i_{max}}  3i_{max}+ \delta_a + \delta_b)  \geq 1-2\delta.
	\]	
	Since the bad events in Eqs.  (\ref{eq:eq11}) and (\ref{eq:eq22}) do not happen,  
	\begin{align}
	\label{eq:num1_1}\hat \I(\hat S_k) &\leq (1+\epsilon_a) \I(\hat S_k),  \\
	\label{eq:num1_3}\hat \I(S^*_k) &\geq (1 - \epsilon_b)\OPT_k, \text{ and }
	\end{align}
	Similar to the proof of Theorem~\ref{theo:type1}, it follows that
	\begin{align}
	\label{eq:num1_2}\I(\hat S_k) &\geq (1-1/e-\epsilon) \OPT_k 
	\end{align}

	\underline{Condition C1}: From the $(1-1/e)$ approximation guarantee of the \textsf{Max-Coverage} algorithm, it follows that 
	\begin{align*}
	\Cov_{\R} (\hat S_k) \geq (1-1/e) \Cov_{\R} (S^*_k)
	\end{align*}
	From Eq. (\ref{eq:num1_3}), 
	\[
	\Cov_{\R}(S^*_k) \geq (1-\epsilon_b)  \frac{\OPT_k}{n} |\R|.
	\]
	Thus, 
	\begin{align*}
	\Cov_{\R}(\hat S_k) &\geq (1-1/e)(1-\epsilon_b) \frac{\OPT_k}{n} |\R|\\
	&\geq (1-1/e)(1-\epsilon)  \frac{\OPT_k}{n} \alpha \Upsilon(\epsilon_0, \frac{\delta}{3i_{max}}) \frac{n}{\OPT_k} \\
	&\geq (1-1/e)(1-\epsilon) \alpha \Upsilon(\epsilon_3, \frac{\delta}{3i_{max}}) \text{ (since } \epsilon_0 \leq \epsilon_3) 					
	\end{align*}	
	Select $\alpha > \frac{(1+\epsilon_1)(1+\epsilon_2)}{(1-1/e)(1-\epsilon)}$, we  have
	\begin{align}
	\label{eq:alphalam}
	\Cov_{\R}(\hat S_k) > \Lambda_1 = (1+\epsilon_1)(1+\epsilon_2)\Upsilon(\epsilon_3, \frac{\delta}{3i_{max}})
	\end{align}	
	
	\underline{Termination of \textsf{Estimate-Inf}}: 	We show that \textsf{Estimate-Inf} does not return $-1$. If \textsf{Estimate-Inf} terminates in line 7, Algo.~\ref{alg:check}, for some $T < T_{max}$, then nothing left to prove. Otherwise, 
	we show that when $|\R_c|=T_{max}$, then \[
	\Cov_{\R_c}(\hat S_k) \geq \Lambda_2 = 1 + (1+\epsilon_2) \Upsilon(\epsilon_2, \frac{\delta}{3i_{max}}) \text{(Line 2, Algo.~\ref{alg:check}),}
	\]
	and, hence, \textsf{Estimate-Inf} returns an estimate but not $-1$.

	By definition of $T_{\PIMA}$, \[|\R| \geq T_{\PIMA}  \geq \alpha\Upsilon(\epsilon_0, \frac{\delta}{3i_{max}}) \frac{n}{\OPT_k} \geq \alpha\Upsilon(\epsilon_2, \frac{\delta}{3i_{max}})\frac{n}{\OPT_k}.\] 
	The last inequality is due to  $\epsilon_0 \leq \epsilon_2$. Thus,
	\begin{align}	
	\label{eq:tmaxb}
	T_{max} = 2|\R|\frac{1+\epsilon_2}{1-\epsilon_2} \frac{\epsilon_3^2}{\epsilon_2^2} 
	\geq 2\alpha\Upsilon(\epsilon_2, \frac{\delta}{3i_{max}}) \frac{\epsilon_3^2}{\epsilon_2^2}\frac{n}{\OPT_k}.
	\end{align}
	Select large enough $\alpha$, says $\alpha > \frac{\epsilon_2^2}{\epsilon_3^2}$, we obtain
	\[
	\epsilon_2^{(i)} = \sqrt{ \frac{(\ln 1/\delta +\ln 3i_{max}) n}{T_{max}} \I(\hat S_k)} \leq \epsilon_2
	\]	
	Since the bad event in Lem.~\ref{lem:eit} does not happen we have
	\begin{align*}
	\I_c(\hat S_k) \geq (1- \epsilon_2^{(i)}) \I(\hat S_k)
	\end{align*}						
	Since $\epsilon_2^{(i)} \leq \epsilon_2$, it follows that
	\begin{align}
	\label{eq:e2c}
	\I_c(\hat S_k) \geq (1- \epsilon_2) \I(\hat S_k)
	\end{align}
	and $\hat \I_c(\hat S_k) = \frac{\Cov_{\R_c}(\hat S_k) n}{|\R_c|}$, we have
	\begin{align*}
	\Cov_{\R_c}(\hat S_k) \geq (1- \epsilon_2) \frac{\I(\hat S_k)}{n} |\R_c| = (1- \epsilon_2) \frac{\I(\hat S_k)}{n} T_{max} 
	\end{align*}
	From Eqs. (\ref{eq:num1_2}) and (\ref{eq:tmaxb}),
	\begin{align}
	\nonumber\Cov_{\R_c}(\hat S_k)& \geq (1-\epsilon_2) \frac{(1-\frac{1}{e}-\epsilon)\OPT_k }{n}
	2\alpha\Upsilon(\epsilon_2, \frac{\delta}{3i_{max}}) \frac{\epsilon_3^2}{\epsilon_2^2}\frac{n}{\OPT_k}\\
	\label{eq:alpha2}&\geq 	1+ (1+\epsilon_2)\Upsilon(\epsilon_2, \frac{\delta}{3i_{max}}).
	\end{align}
	Here, we select $\alpha > 1+ \frac{\epsilon_2^2}{\epsilon_3^2}\frac{1}{2 (1-\epsilon_2)(1-1/e-\epsilon)}$.
	
	\underline{Condition C2}: We show that the condition C2, $\hat \I(\hat S_k)  \leq (1+\epsilon_1) I_c(\hat S_k)$, in line 11, Algo.~\ref{alg:main} is satisfied with proper selection of $\epsilon_1, \epsilon_2, \epsilon_3$.	The condition C2 is equivalent to 
	\[
	\frac{\hat \I(\hat S_k)}{\I_c(\hat S_k)} - 1 \leq \epsilon_1 \Leftrightarrow 		\frac{\hat \I(\hat S_k)}{\I(\hat S_k)}  \frac{\I(\hat S_k)}{\I_c(\hat S_k)} - 1 \leq \epsilon_1
	\]
	From Eqs. (\ref{eq:num1_1}) and (\ref{eq:e2c}), we have
	\[
	\frac{\hat \I(\hat S_k)}{\I(\hat S_k)}  \frac{\I(\hat S_k)}{\I_c(\hat S_k)} - 1 \leq (1+\epsilon_a)\frac{1}{1-\epsilon_2} -1 = \frac{\epsilon_a + \epsilon_2}{1-\epsilon_2} 
	\]	
	Set $\epsilon_1 = \frac{\epsilon_a +\epsilon_b/2}{1-\epsilon_b/2}, \epsilon_2 = \epsilon_b/2, \epsilon_3 = \frac{\epsilon_b^2}{2 - \epsilon_b}$, the following holds

	\begin{align}
	\label{eq:setting123}
	\begin{cases} 
	\epsilon_1 \in (0, \infty), \epsilon_2, \epsilon_3 \in (0, 1) \\			
	(1-\frac{1}{e})\frac{\epsilon_a + \epsilon_b}{1+\epsilon_a} =
	(1-\frac{1}{e})\frac{\epsilon_1+\epsilon_2+\epsilon_1\epsilon_2+\epsilon_3}{(1+\epsilon_1)(1+\epsilon_2)} = \epsilon\\ \frac{\hat \I(\hat S_k)}{\I_c(\hat S_k)} - 1 \leq  \frac{\epsilon_a + \epsilon_2}{1-\epsilon_2} = \epsilon_1			
	\end{cases} 
	\end{align}
	Thus, \PIMA{} with the setting in (\ref{eq:setting123}) will stop with a probability at least $1-2\delta$ if $|\R| \geq T_{\PIMA} = O( N_1(\epsilon_a, \epsilon_b, \delta_a, \delta_b) )$.
	
	\underline{Constants  Justification}: The factors that are assumed to be constants within our proofs for \PIMA{} are 
	1) the factor  $3\frac{\epsilon_b^2}{\epsilon_0}$ in  Eq.~(\ref{eq:tssac}), 
	2) $\alpha > \frac{(1+\epsilon_1)(1+\epsilon_2)}{(1-1/e)(1-\epsilon)}$  before Eq.~(\ref{eq:alphalam}),
	3)  $\alpha > \frac{\epsilon_2^2}{\epsilon_3^2}$,  after Eq.~(\ref{eq:tmaxb}), 
	4) $\alpha > 1+ \frac{\epsilon_2^2}{\epsilon_3^2}\frac{1}{2 (1-\epsilon_2)(1-1/e-\epsilon)}$,  after Eq.~(\ref{eq:alpha2}).
	With the above setting of $\epsilon_1, \epsilon_2, \epsilon_3$, we can verify that those factors are constants that depend only on $\epsilon, \epsilon_a$, and $\epsilon_b$.
	
\subsection*{Proof of Lemma~\ref{lem:tmax}}
	\begin{align}
	t_{\max} & = \log_2 (\frac{2 N_{\max}}{\Upsilon(\epsilon, \delta/3)}) \nonumber \\
	& = \log_2 \left (2(2-\frac{1}{e})^2\frac{(2+\frac{2}{3}\epsilon)n \cdot \frac{\ln (6/\delta)+\ln {n\choose k}}{k\epsilon^2}}{(2+\frac{2}{3}\epsilon)\ln (\frac{3}{\delta}) \frac{1}{\epsilon^2}} \right) \nonumber \\
	& = \log_2 \left(2(2-\frac{1}{e})^2 \frac{n (\ln (6/\delta)+ \ln{n \choose k})}{k \ln (3/\delta)} \right) \nonumber \\
	& \leq \log_2 \left(2 (2-\frac{1}{e})^2\frac{n (\ln (6/\delta)+ k \ln n)}{k \ln (3/\delta)} \right) \nonumber \\
	& \leq 2\log_2 n + 2 = O(\log_2 n)
	\end{align}
	The last inequality follows from our assumption $1/\delta = \Omega(\log_2 n)$.
	
\subsection*{Proof of Lemma~\ref{lem:bad2}}
	One can verify that $f(x)$ is a strictly decreasing function for $x>0$. Moreover, $f(0)=1$ and $\lim_{x\rightarrow \infty} f(x) = 0$. Thus, the equation $f(x)=\frac{\delta}{3t_{max}}$ has an {\em unique solution} for $0<\delta<1$ and $t_{max}\geq 1$.
	
	{\em Bound the probability of $B^{(2)}_t$}: 
	Note that $\hat \epsilon_t$ and the samples generated in $\R^{c}_t$ are independent. Thus, we can apply the concentration inequality in Eq.~(\ref{eq:plus}):
	\begin{align*}
	&\Pr[\I^{c}_t(\hat S_k) \geq (1 + \hat \epsilon_t)\I(\hat S_k)]
	\leq \exp\left({-\frac{N_t \I(\hat S_k) {\hat \epsilon_t}^2}{(2+\frac{2}{3}\hat \epsilon_t) n}}\right)  =\frac{\delta}{3 t_{max}}.
	\end{align*}
	The last equation is due to the definition of $\hat \epsilon_t$.	
	
	{\em Bound the probability of $B^{(3)}_t$}: Since $\epsilon^*_t$ is fixed and independent from the generated samples, we have
	\begin{align}
	\Pr[& \hat \I_t(S^*_k) \leq (1-\epsilon^*_t) \OPT_k]  \leq \exp\left({- \frac{|\R_t| \OPT_k {\epsilon^*_t}^2}{2 n}}\right) \nonumber \\
	& = \exp\left({-\frac{\Lambda 2^{t-1} \OPT_k \epsilon^2 n }{2 n 2^{t-1} \OPT_k}}\right) \\
	&= 		\exp\left({- \frac{(2+\frac{2}{3}\epsilon) \ln (\frac{3 t_{max}}{\delta})\frac{1}{\epsilon^2} 2^{t-1} \OPT_k \epsilon^2 n}{ 2(1+\epsilon/3)n 2^{t-1} \OPT_k}}\right) \nonumber \\
	& \leq \exp\left({-\ln \frac{3 t_{max}}{\delta}}\right) = \frac{\delta}{3 t_{max}},
	\end{align}
	which completes the proof of Lemma.~\ref{lem:bad2}.

\subsection*{Proof of Lemma~\ref{lem:e2e}}
	Since the bad event $B^{(2)}_t$ doesn't happen, we have
	\begin{align}
	\nonumber \hat \I_t^{(c)}(\hat S_k) \leq (1+\hat \epsilon_t)\I(\hat S_k)
	\Rightarrow \Cov_{\R^{c}_t}(\hat S_k) \leq (1+\hat \epsilon_t)N_t\frac{\I(\hat S_k)}{n}
	\end{align}
	
	When \DPIMA{} stops with $\epsilon_t \leq \epsilon$, it must satisfy the condition on line 9 of \DPIMA{}
	\[
	\Cov_{\R^{c}_t}(\hat S_k) \geq \Lambda_1.
	\] 
	Thus, we have
	\begin{align}
	\label{eq:Nt}(1+\hat \epsilon_t)N_t\frac{\I(\hat S_k)}{n} \geq \Lambda_1=1+(1+\epsilon)\frac{2+2/3\epsilon}{\epsilon^2} \ln \frac{3t_{max}}{\delta}
	\end{align}
	From the definition of $\hat \epsilon_t$, it follows that
	\begin{align}
	\label{eq:nt}
	N_t = \frac{2+2/3 \hat \epsilon_t}{\hat\epsilon_t^2} \ln \left(\frac{3t_{max}}{\delta}\right)\frac{n}{\I(\hat S_k)}
	\end{align}
	Substitute the above into (\ref{eq:Nt}) and simplify, we obtain:
	\begin{align}
	&(1+\hat \epsilon_t)\frac{2+2/3 \hat \epsilon_t}{\hat\epsilon_t^2} \ln \left(\frac{3t_{max}}{\delta}\right)\\
	\geq & (1+\epsilon)\frac{2+2/3\epsilon}{\epsilon^2} \ln \frac{3t_{max}}{\delta}+1
	\end{align}
	Since the function $(1+x)\frac{2+2/3x}{x^2}$ is a decreasing function for $x>0$, it follows that $\hat \epsilon_t < \epsilon$.

\subsection*{Proof of Theorem~\ref{theo:accuracydssa}}

	
	Assume that none of the bad events $B^{(1)}, B^{(2)}_t, B^{(3)}_t$  ($t =1..t_{max}$) in Lemmas \ref{lem:cap} and \ref{lem:bad2} happens. Apply union bound, the probability that the assumption holds is at least
	\begin{align}
	1-(\delta/3 + \left(\delta/(3t_{max}) + \delta/(3t_{max})\right) \times t_{max}) \geq 1 -\delta
	\end{align}
	
	We shall show that the returned $\hat S_k$ is a $(1-1/e-\epsilon)$-approximation solution. If \DPIMA{} stops with $|\R_t| \geq N_{max}$, $\hat S_k$ is a $(1-1/e-\epsilon)$-approximation solution, since the bad event $B^{(1)}$ does not happen.
	
	Otherwise, \DPIMA{} stops at some iteration $t$ and $\epsilon_t \leq \epsilon$. We use contradiction method.  
	Assume that 
	\begin{align}
	\label{eq:contrad}
	\I(\hat S_k) < (1-1/e-\epsilon) \OPT_k.
	\end{align}
	The proof will continue in the following order
	\begin{enumerate}[label=(\Alph*)]
		\item  $\I(\hat S_k) \geq (1-1/e-\epsilon_t') \OPT_k$\\ where $\epsilon'_t = (\epsilon_1 + \hat\epsilon_t + \epsilon_1 \hat\epsilon_t)(1-1/e-\epsilon) + (1-1/e)\epsilon^*_t$.
		\item  $\hat \epsilon_t \leq \epsilon_2$ and  $\epsilon^*_t \leq \epsilon_3$.
		\item  $\epsilon_t' \leq \epsilon_t \leq \epsilon \Rightarrow \I(\hat S_k) \geq (1-\frac{1}{e}-\epsilon) \OPT_k$ (\emph{contradiction}).
	\end{enumerate}

	\emph{Proof of (A)}. Since the bad events $B^{(2)}_t$ and $B^{(3)}_t$ do not happen, we have
	\begin{align}
	\label{eq:ici} \hat \I_t^{(c)}(\hat S_k) &\leq (1+\hat \epsilon_t)\I(\hat S_k), \text{and}\\
	\label{eq:iho} \hat \I_t(S^*_k) &\leq (1-\epsilon_t^*) \OPT_k.
	\end{align}
	Since $\epsilon_1 \leftarrow \hat \I_t(\hat S_k)/\I^{c}_t(\hat S_k) - 1$, it follows from (\ref{eq:ici}) that
	\begin{align}
	\hat \I_t(\hat S_k) =  (1+\epsilon_1)\I^{c}_t(\hat S_k)\leq (1+\epsilon_1)(1 + \hat\epsilon_t) \I(\hat S_k) \nonumber
	\end{align}	
	Expand the right hand side and apply (\ref{eq:contrad}), we obtain	
	\begin{align}
	\I(\hat S_k) &\geq \hat \I_t(\hat S_k) - (\epsilon_1 + \hat\epsilon_t + \epsilon_1 \hat\epsilon_t) \I(\hat S_k) \nonumber \\
	& \geq \hat \I_t(\hat S_k) - (\epsilon_1 + \hat\epsilon_t + \epsilon_1 \hat\epsilon_t) (1-1/e-\epsilon)\OPT_k \nonumber
	\end{align}
	Since the greedy algorithm in the Max-Coverage guarantees a $(1-1/e)$-approximation, $\hat \I_t(\hat S_k) \geq (1-1/e) \hat \I_t(S^*_k)$. Thus
	\begin{align}		
	&\I(\hat S_k)  \geq (1-1/e) \hat \I_t(S^*_k)  - (\epsilon_1 + \hat\epsilon_t + \epsilon_1 \hat\epsilon_t)(1-1/e-\epsilon) \OPT_k \nonumber \\
	&\geq (1-1/e)(1-\epsilon^*_t) \OPT_k  - (\epsilon_1 + \hat\epsilon_t + \epsilon_1 \hat\epsilon_t)(1-1/e-\epsilon) \OPT_k  \nonumber \\
	&\geq (1-1/e - \epsilon'_t) \OPT_k, \nonumber
	\end{align}
	where $\epsilon'_t = (\epsilon_1 + \hat\epsilon_t + \epsilon_1 \hat\epsilon_t)(1-1/e-\epsilon) + (1-1/e)\epsilon^*_t$.
	
	\emph{Proof of (B)}. We show that $\hat \epsilon_t \leq \epsilon_2$. Since, $\epsilon_2 = \epsilon \sqrt{\frac{n(1+\epsilon)}{2^{t-1} \I^{c}_t (\hat S_k)}}$, we have
	\[
	\frac{1}{\epsilon^2} = \frac{1}{\epsilon_2^2} \frac{n}{2^{t-1}} \frac{1+\epsilon}{\I^{c}_t(\hat S_k)}.
	\]
	Expand the number of RR sets in iteration $t$, $N_t = 2^{t-1} \Lambda$, and apply the above equality, we have 
	\begin{align}
	N_t &=  2^{t-1} (2+2/3\epsilon)\frac{1}{\epsilon^2} \ln \frac{3t_{max}}{\delta} \\
	&=  2^{t-1} (2+2/3\epsilon)\frac{1}{\epsilon_2^2}  \frac{n}{2^{t-1}} \frac{1+\epsilon}{\I^{c}_t(\hat S_k)}  \ln \frac{3t_{max}}{\delta} \\
	&= (2+2/3\epsilon)\frac{1}{\epsilon_2^2}   \frac{(1+\epsilon)n}{\I^{c}_t(\hat S_k)}  \ln \frac{3t_{max}}{\delta}
	\end{align}
	On the other hand, according to Eq.~(\ref{eq:nt}), we also have,
	\begin{align}
	N_t = \frac{2+2/3 \hat \epsilon_t}{\hat\epsilon_t^2} \ln \left(\frac{3t_{max}}{\delta}\right)\frac{n}{\I(\hat S_k)}.
	\end{align}
	Thus
	\begin{align*}
	& (2+2/3\epsilon)\frac{1}{\epsilon_2^2}   \frac{1+\epsilon}{\I^{c}_t(\hat S_k)} = \frac{2+2/3 \hat \epsilon_t}{\hat\epsilon_t^2} \frac{1}{\I(\hat S_k)}\\
	\Rightarrow &\frac{\hat \epsilon_t^2}{\epsilon_2^2} = \frac{2+2/3 \hat \epsilon_t}{2+2/3 \epsilon}  \frac{\I^{c}_t(\hat S_k)}{(1+\epsilon)\I(\hat S_k)} \leq 1
	\end{align*}
	The last step is due to Lemma \ref{lem:e2e}, i.e., $\I^c_t(\hat S_k) \leq (1+\epsilon)\I(\hat S_k)$ and $\hat \epsilon_t \leq \epsilon$. Therefore, $\hat\epsilon_t \leq \epsilon_2$.
	
	We show that $\epsilon^*_t \leq \epsilon_3$. According to the definition of $\epsilon^*_t$ and $\epsilon_3$, we have
	\begin{align*}
	\frac{(\epsilon^*_t)^2}{\epsilon_3^2} &=\frac{n}{(1+\epsilon/3)2^{t-1} \OPT_k} / 
	\frac{n(1+\epsilon)(1-1/e-\epsilon)}{(1+\epsilon/3)2^{t-1} \I^{c}_t (\hat S_k)}\\
	=& \frac{\I^{c}_t (\hat S_k)}{\OPT_k (1+\epsilon)(1-1/e-\epsilon)}
	\leq \frac{\I_t (\hat S_k)}{\OPT_k (1-1/e-\epsilon)}\leq 1
	\end{align*}
	The last two steps follow from Lem. \ref{lem:e2e}, $\I^c_t(\hat S_k) \leq (1+\epsilon)\I(\hat S_k)$ and the assumption 
	(\ref{eq:contrad}), respectively. Thus, $\epsilon_t^* \leq \epsilon_3$.
	
	\emph{Proof of (C)}. Since $1+\epsilon_1 = \hat \I_t(\hat S_k)/\I^{c}_t(\hat S_k) \geq 0$ and $\epsilon_2 \geq \hat \epsilon_t >0$ and 
	$\epsilon_3 \geq \epsilon^*_t >0$, we have
	\begin{align}
	\epsilon_t' &= (\epsilon_1 + \hat\epsilon_t + \epsilon_1 \hat\epsilon_t)(1-1/e-\epsilon) + (1-1/e)\epsilon^*_t \\
	&= (\epsilon_1 + \hat\epsilon_t (1+\epsilon_1))(1-1/e-\epsilon) + (1-1/e)\epsilon^*_t \\
	&\leq (\epsilon_1 + \epsilon_2 (1+\epsilon_1))(1-1/e-\epsilon) + (1-1/e)\epsilon_3  \\
	&= \epsilon_t \leq \epsilon.
	\end{align}

\subsection*{Proof of Theorem~\ref{theo:dima}}

	
	Since $T_2 \geq 1$, there exist $\epsilon_a^*, \epsilon_b^*, \delta_a^*, \delta_b^*$ that satisfy
	\begin{align}
	\label{eq:risp3}
	N^{(1)}_{min}(\epsilon^*_a, \epsilon^*_b, \delta^*_a, \delta^*_b)&=T_2, \\
	\label{eq:risp4}
	(1-\frac{1}{e})\frac{\epsilon_a^*+ \epsilon_b^*}{1+\epsilon_a^*}&= \epsilon \leq \frac{1}{4}, \\
	\label{eq:risp5}
	\delta_a^* + \delta_b^* &\leq \delta <\frac{1}{\log_2 n}.
	\end{align}
	
	Let  $\epsilon_0 = \min\{ \epsilon, \epsilon^*_b\}$, and 
	\begin{align}
	T_{\DPIMA} = \max\{ T_2, \alpha \Upsilon(\epsilon_0, \frac{\delta}{3t_{max}}) \frac{n}{\OPT_k}\},
	\end{align}
	for some constant $\alpha$ specified later. Note that $\frac{\epsilon^*_b}{\epsilon} \leq 1/(1-1/e)$. Similar to the proof in Lem.~\ref{lem:tssa}, we can show that \[T_{\DPIMA} = O( T_2)\] under the \emph{range conditions}.

	From Def.~\ref{def:type1_min} of the type-1 minimum threshold,   if $|\R|\geq T_{\DPIMA} \geq T_2$ then
	\begin{align}
	\label{eq:eq1}
	&\Pr [\hat \I(\hat S_k) > (1 + \epsilon_a^*)\I(\hat S_k)] \leq \delta_a^* \text{ and }\\
	\label{eq:eq2}
	&\Pr [\hat \I(S^*_k) < (1 - \epsilon_b^*)\OPT_k] \leq \delta_b^*.
	\end{align}
	Similar to the proof of Theorem~\ref{theo:type1}, it follows that
	\begin{align}
	\label{eq:num2_2}\I(\hat S_k) &\geq (1-1/e-\epsilon) \OPT_k 
	\end{align}

	Assume that \DPIMA{} reaches to a round $t \leq t_{max}$ with  $|\R| = \Lambda 2^{t-1} \geq T_{\DPIMA}$. If $T_{\DPIMA} > N_{max}$ then \DPIMA{} will stop and the proof is complete. Otherwise consider the bad events that $\I_t^c(\hat S_k) = \frac{\Cov_{\R^{c}_t}(\hat S_k)n}{|\R^{c}_t|}$ is an underestimate of $\I(\hat S_k)$. Specifically, define for each $t=1,\ldots,t_{max}$ the event
	\[
	B^{(4)}_t = \left( \hat \I^c_t(\hat S_k) < (1-\tilde\epsilon_t)\I(\hat S_k)\right),
	\]
	where $\tilde{\epsilon_t} = \epsilon \sqrt{\frac{n}{(1+\epsilon/3)2^{t-1} \I(\hat S_k)}}$. Similar to the proof of Lem.~\ref{lem:bad2}, we can show that \[
	\Pr[B^{(4)}_t] \leq \frac{\delta}{3t_{max}}.
	\]
	
	Assume that neither  the bad events in Eqs. (\ref{eq:eq1}) and (\ref{eq:eq2}) nor the bad events $B^{(2)}_t,B^{(3)}_t, B^{(4)}_t$  happen for any $t\in [1, t_{max}]$. Apply the union bound, this assumption holds with a probability at least 
	\[
	1- (\delta^*_a + 	\delta^*_b +  \frac{\delta}{3t_{max}} t_{max} + \frac{\delta}{3t_{max}} t_{max} + \frac{\delta}{3t_{max}} t_{max}) \geq 1 - 2 \delta.
	\]
	Under the above assumption, we will show that the two conditions D1 and D2 are met, and, thus, \DPIMA{} will stop.
	
	\underline{Condition D1}: We will prove that $\Cov_{\R^{c}_t}(\hat S_k) \geq \Lambda_1$. 
	Since $|\R| \geq T_{\DPIMA} \geq \alpha\Upsilon(\epsilon_0, \frac{\delta}{3t_{max}} ) \frac{n}{\OPT_k}$ and $\epsilon_0 \leq \epsilon$, we have
	\begin{align}
	\nonumber 2^{t-1} &\geq \frac{|\R|}{\Upsilon(\epsilon, \frac{\delta}{3t_{max}})} 
	\geq \alpha\Upsilon(\epsilon_0, \frac{\delta}{3t_{max}} ) \frac{n}{\OPT_k}/ \Upsilon(\epsilon, \frac{\delta}{3t_{max}})\\
	\label{eq:2t1}&\geq \alpha \frac{n}{\OPT_k} \frac{\epsilon^2}{\epsilon_0^2}
	\end{align}
	
	Select $\alpha > 9/(1-1/e-\epsilon)$ and apply Eq. \ref{eq:num2_2}, we have
	\begin{align}
	\nonumber\tilde{\epsilon_t} &= \epsilon \sqrt{\frac{n}{(1+\epsilon/3)2^{t-1} \I(\hat S_k)}} \\
	\label{eq:etd}&\leq \epsilon \sqrt{\frac{n}{(1+\epsilon/3)\alpha \frac{n}{\OPT_k}  \frac{\epsilon^2}{\epsilon_0^2} (1-1/e-\epsilon) \OPT_k}} \leq \frac{\epsilon_0}{3}
	\end{align}

	Since  $B^{(4)}_t$ does not happen, we have
	\begin{align}
	\label{eq:etc}\hat \I^c_t(\hat S_k) \geq (1-\tilde\epsilon_t)\I(\hat S_k).
	\end{align}
	We have
	\begin{align*}
	&\Cov_{\R^{c}_t}(\hat S_k)=\frac{\Cov_{\R^{c}_t}(\hat S_k) n}{|R^{c}_t|} \frac{|R^{c}_t|}{n} =\hat \I^c_t(\hat S_k) \frac{|\R^{c}_t|}{n} \\
	&\geq (1 - \tilde\epsilon_t)  \frac{\I(\hat S_k)}{n}  \Upsilon(\epsilon, \frac{\delta}{3 t_{max}}) 2^{t-1}\\
	&\geq (1-\epsilon/3)(1-1/e-\epsilon) \frac{\OPT_k}{n}  \Upsilon(\epsilon,\frac{\delta}{3t_{max}})  \alpha\frac{n}{\OPT_k}\\
	&\geq 1+(1+\epsilon) \Upsilon(\epsilon, \frac{\delta}{3 t_{max}}) = \Lambda_1,
	\end{align*}
	when selecting $\alpha > 1 + \frac{1}{(1-1/e-\epsilon)(1-\epsilon/3)}$.
	
	\underline{Condition D2}: $\epsilon_t \leq \epsilon$.
	The condition D2 is equivalent to 
	\begin{align}
	\nonumber  &(1-1/e)\frac{\epsilon_1+\epsilon_2+\epsilon_1 \epsilon_2 +\epsilon_3}{(1+\epsilon_1)(1+\epsilon_2)} \leq \epsilon = (1-1/e) \frac{\epsilon^*_a + \epsilon^*_b}{1+\epsilon^*_a} \\
	\nonumber\Leftrightarrow &1-\frac{1-\epsilon_3}{(1+\epsilon_1)(1+\epsilon_2)} \leq 1-\frac{1-\epsilon^*_b}{1+\epsilon^*_a} \\
	\Leftrightarrow &1-\epsilon_3 \geq \frac{1-\epsilon^*_b}{1+\epsilon^*_a}(1+\epsilon_1)(1+\epsilon_2) 
	\end{align}
	From Eqs. (\ref{eq:eq1}), (\ref{eq:etc}),  and (\ref{eq:etd}), we have
	\[
	1 + \epsilon_1 = 	\frac{\hat \I(\hat S_k)}{\I(\hat S_k)}  \frac{\I(\hat S_k)}{\I^c_t(\hat S_k)} \leq (1+\epsilon^*_a)\frac{1}{1-\tilde\epsilon_t} \leq  \frac{1+\epsilon^*_a}{1-\epsilon_0/3} \leq \frac{1+\epsilon^*_a}{1-\epsilon^*_b/3}
	\]
	Thus, it is sufficient to show that
	\begin{align}
	\nonumber&1-\epsilon_3 \geq  \frac{1-\epsilon^*_b}{1+\epsilon^*_a} \frac{1+\epsilon^*_a}{1-\epsilon^*_b/3} (1+\epsilon_2)\\
	\nonumber\Leftrightarrow &(1-\epsilon_3)(1-\epsilon^*_b/3) \geq (1-\epsilon^*_b)(1+\epsilon_2)\\
	\label{eq:lasteq}\Leftrightarrow &\frac{2}{3}\epsilon^*_b+ \frac{\epsilon^*_b}{3}\epsilon_3 + \epsilon^*_b \epsilon_2 \geq \epsilon_2+\epsilon_3
	\end{align}

	Apply the inequalities $2^{t-1} \geq \alpha \frac{n}{\OPT_k}$, Eq. (\ref{eq:2t1}), and $\I(\hat S_k) \geq (1-1/e-\epsilon) \OPT_k$, Eq. (\ref{eq:num2_2}). For sufficiently large $\alpha>  \frac{9(1+\epsilon)}{(1-1/e-\epsilon)}$, we have
	\begin{align}
	\epsilon_2 &= \epsilon \sqrt{\frac{n(1+\epsilon)}{2^{t-1} \I^{c}_t (\hat S_k)}} \leq \epsilon_0/3 \leq \epsilon^*_b/3\\
	\epsilon_3 &= \epsilon \sqrt{\frac{n(1+\epsilon)(1-1/e-\epsilon)}{(1+\epsilon/3)2^{t-1} \I^{c}_t (\hat S_k)}}	\leq \epsilon_0/3 \leq \epsilon^*_b/3
	\end{align}
	Therefore, $\epsilon_2+\epsilon_3 \leq 2/3 \epsilon^*_b$ and the inequality (\ref{eq:lasteq}) holds.
	This completes the proof.

\end{document}